%% file: main.tex
\pdfoutput=1
\documentclass[%
 reprint,
 amsmath,amssymb,
 aps,
letterpaper,
]{revtex4-2}

\usepackage{graphicx}
\usepackage{dcolumn}
\usepackage{bm}

\usepackage[english]{babel}
\usepackage[T1]{fontenc}
\usepackage[version=4]{mhchem}
\usepackage{balance}
\usepackage[caption=false]{subfig}
\usepackage{placeins}

\usepackage[hyperindex,breaklinks]{hyperref}  
\begin{document}

\preprint{APS/123-QED}

\title{Unsupervised learning of representative local atomic arrangements\\ in molecular dynamics data}

\author{Fabrice Roncoroni}
\author{Ana Sanz-Matias}%
\author{Siddharth Sundararaman}
\author{David Prendergast}
    \email{dgprendergast@lbl.gov}
 
\affiliation{%
    Joint Center for Energy Storage Research, the Molecular Foundry, Lawrence Berkeley
    National Laboratory, Berkeley, California 94720, United
    States
}%

\date{\today}

\begin{abstract}
\input{Sections/00_abstract.tex}
\end{abstract}

\maketitle

\input{Sections/01_introduction}
\input{Sections/02_methodology}

\input{Sections/03_computational_details}

\input{Sections/04_results}

\input{Sections/05_outlook_discussion.tex}
\input{Sections/06_conclusions}

\FloatBarrier

\section*{Author Contributions}
All authors helped conceptualize the original idea from DP. ASM and FR developed and tested the clustering algorithm. FR ran the MD calculations, analyzed the data and wrote the original manuscript draft with support from ASM and DP. SS contributed to force field development. DP supervised and managed the project.

\section*{Conflicts of interest}
There are no conflicts to declare.

\section*{Acknowledgements}
This work was fully supported by the Joint Center for Energy Storage Research (JCESR), an Energy Innovation Hub funded by the U.S. Department of Energy, Office of Science, Basic Energy Sciences. All simulations were performed at the Molecular Foundry, Lawrence Berkeley National Laboratory.



\balance

\bibliography{bibliography}
\bibliographystyle{bibstyle}

\end{document}

%% file: Sections/00_abstract.tex
\section*{Abstract}
Molecular dynamics (MD) simulations present a data-mining challenge, given that they can generate a considerable amount of data but often rely on limited or biased human interpretation to examine their information content. By not asking the right questions of MD data we may miss critical information hidden within it. We combine dimensionality reduction (UMAP) and unsupervised hierarchical clustering (HDBSCAN) to quantitatively characterize prevalent coordination environments of chemical species within MD data. By focusing on local coordination, we significantly reduce the amount of data to be analyzed by extracting all distinct molecular formulas within a given coordination sphere. We then efficiently combine UMAP and HDBSCAN with  alignment or shape-matching algorithms to partition these formulas into structural isomer families indicating their relative populations. The method was employed to reveal details of cation coordination in electrolytes based on molecular liquids.


%% file: Sections/01_introduction.tex
\section*{Introduction}

Molecular dynamics (MD) is a vital tool in gaining molecular-scale insight on the properties and functional behavior of complex systems and interfaces.
There is immense value and inspiration in providing the community with visualizations of molecular configurations and their dynamics, particularly with respect to the identification of molecular-scale bottlenecks in functional processes and the rational design of new chemistries to avoid them \cite{VV2014, Sun2020, Yang2020, Yamijala2021, Young2021, Yao2022}.

Analysis of MD data sets is frequently driven by simplistic metrics (e.g., density profiles, pair distribution functions) and human intuition on what questions to ask of the data. Insight can be limited significantly by human imagination or past experience when it comes to designing bespoke analysis of such data sets. The underlying assumption, which we presume here also, is that the relevant information content in large data sets has a much smaller dimension than the entire data set, or that there is a relatively simple, low-dimensional underlying probability density that can describe behavior in the system. In an ideal scenario, automated data analysis should provide us with an unbiased path to dimensionality reduction with full disclosure of embedded details of the distribution/density in terms of distinct classification or grouping of data, while the task of labeling these groups (via interpretation or contextualization) is more suited to human intuition and experience.

To this end, we have developed an unsupervised data-mining approach to extract details of distinct motifs of local coordination from MD simulations. We divide the problem of assessing prevalent local coordination environments into three main steps. First, we identify the species of interest and the expected size of their local coordination environment. Then, we extract local atomic clusters from the MD data set and group them by chemical formula or composition. Finally for each chemical formula, we perform dimensionality reduction, hierarchical clustering and alignment procedures to determine the structural distribution of coordination isomers. We focus on defining a computationally efficient approach, which can quickly analyze large MD data sets to provide details of conformations within minutes or less to facilitate on-the-fly analysis. To this end, the combination of efficient dimensionality reduction (UMAP~\cite{McInnes2018}) with hierarchical clustering (HDBSCAN~\cite{McInnes2017}) is aided significantly by our use of fast alignment algorithms (FASTOVERLAP \cite{Griffiths2017}) that focus on relative alignment with respect to automatically identified exemplars of isolated data clusters. Furthermore, we demonstrate that this approach can be used to "learn" how to classify previously unseen data and perform "on-the-fly" unsupervised analysis of MD data as it is generated. The outcome of this approach is a detailed molecular-scale understanding of the local coordination environment of particular species, with a significant reduction in data dimensionality, that is informed more by the data itself than by human expectation (i.e., minimizing external bias).

In what follows, we provide a case study for mining the coordination environments of ions in liquid electrolytes as provided by extensive molecular dynamics sampling using empirical force fields, computed using LAMMPS \cite{LAMMPS}. The sampling may result from simple trajectories for well-defined thermodynamic ensembles, or it may be biased with respect to certain collective variables, as in umbrella sampling \cite{Torrie1977} or metadynamics \cite{Laio2002, Fiorin2013, Bussi2020}, to explore the free energy landscape within a defined low-dimensional collective variable space.

Our chosen example focuses on multivalent ions in nonaqueous electrolytes as a particularly challenging case. In our experience, the distribution of coordination environments about multivalent ions is definitively multimodal -- one should not speak of \emph{the} solvation environment, but rather admit that there may be multiple solvation environments \cite{Baskin2019}. This presents several challenges to theoretical modeling, especially when particular ion coordinations may persist in deep local minima of free energy, hindering a full sampling of the available configuration space. Although such challenges have been addressed using collective variables that span coordination number, \cite{Roy2016, Byrne2017} we may lack validation that the sampled behavior is not limited by the choice of collective variable. In addition, experimental characterization, such as spectroscopy, requires molecular models for interpretation, especially if multiple isomers may contribute to the measured spectra \cite{Camacho2015, Nandasiri2017, Hu2018, Yang2020, Hahn2020, Kao2020, Agarwal2021}. For future work on this complex problem, we claim that unsupervised classification can ultimately deconvolute the distribution of coordination environments into well-separated unimodal, normal distributions, which can be provided as motifs for defining optimal collective variables to sample the free energy landscape effectively and to make unambiguous interpretations of experimental measurements.

%% file: Sections/02_methodology.tex
\section*{Methodology}

The behavior of electrolyte, in bulk or in the presence of an interface, can be studied with MD simulations~\cite{Xu2004, Kerisit2006, Bedrov2019}. Because of the high amount of data generated even by a simple MD trajectory, the task to analyze and identify relevant aspects of the calculation can become cumbersome. In the field of electrolyte science, often we are interested in the unique chemical environments that are present around specific solvated ions. In particular, the first solvation shell is critical in determining important properties, such as solubility of the salt, transport properties, stability and reactivity with the electrodes~\cite{Pham2017, Rajput2018, Han2019, Bezabh2020, Hou2021, Tian2022, Konstantinovsky2022, Yu2022}. In the following section we present a workflow we developed to assist with the analysis of large MD datasets. Our goal is to understand which solvation environments are sampled by the MD trajectory. Ideally, we want to minimally rely on arbitrary inputs provided by a user and instead make use of a robust, unbiased procedure that can leverage as much data as is available to us.

\subsection*{Local structure sampling}

Since we are interested in the chemical environment around solvated ions in electrolytes, the first step is to define a spherical region around the ions of interest (e.g. the \ce{Ca^{2+}} cation) with an element-specific cut off radius and remove the atoms outside it. In this way, we can effectively isolate what we refer in this paper as the \emph{local atomic arrangement} around the ion of interest. An effective choice to define the cut off is to use element-specific radial distribution functions and estimate the extent of the first solvation shell. Extracting the local atomic arrangement out of an entire MD frame dramatically reduces the size of the trajectory to be analyzed. Instead of frames of thousands of atoms we are left with a small cluster of atoms around the species of interest. If there is more than one ion of interest in the simulation box, multiple local atomic arrangements can exist in each frame. In cases where their solvation shells may overlap, we can further decide to merge them into dimers, trimers, etc. and consider them as a single unit.

\begin{figure*}[ht]
    \centering
    \includegraphics[width=\linewidth]{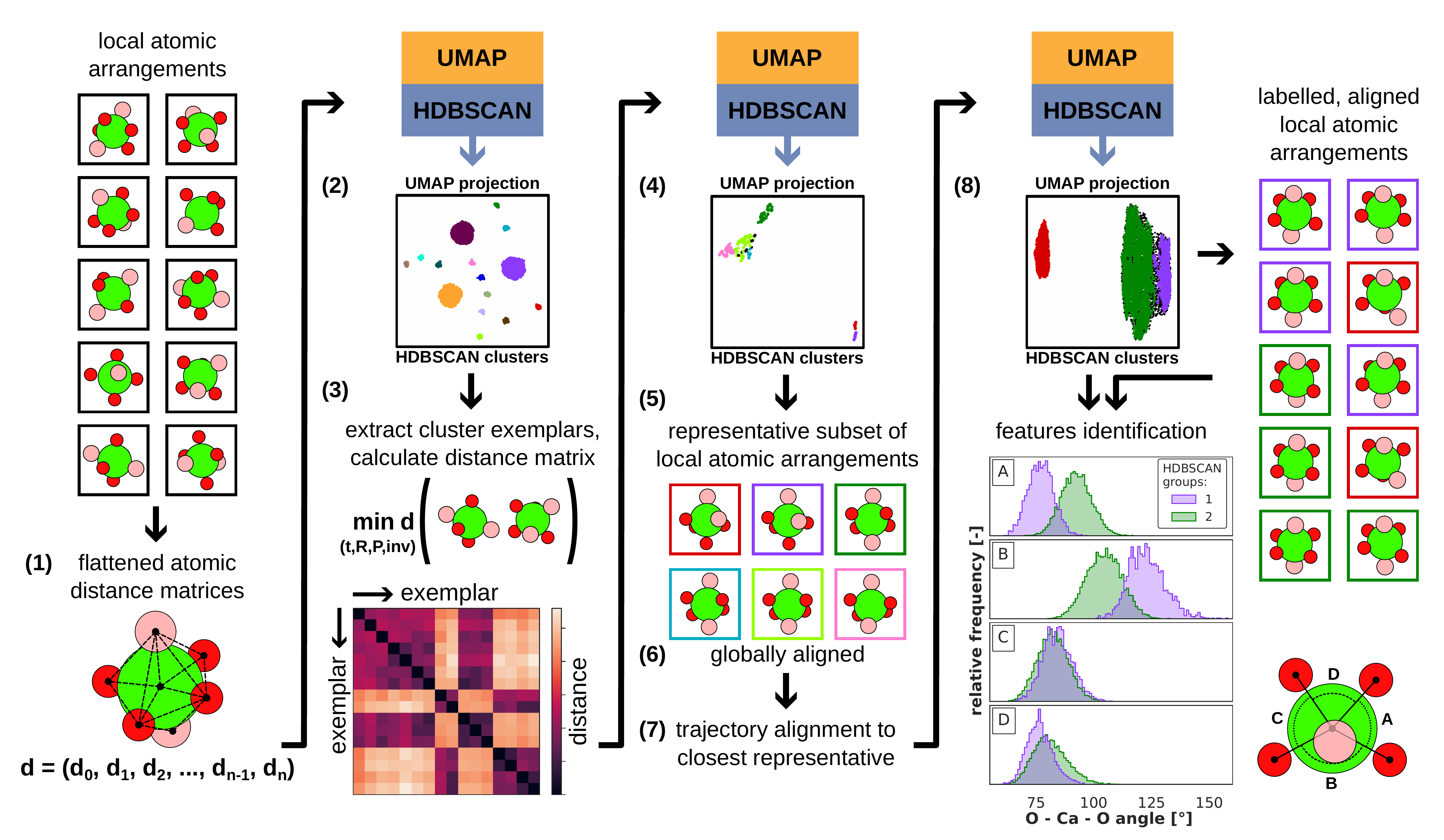}
    \caption{Schematic described in the text highlighting the multi-step process by which local atomic arrangements for a given chemical formula are ultimately grouped into clusters of aligned atomic arrangements through successive application of dimensionality reduction (UMAP), hierarchical density-based clustering (HDBSCAN) and alignment via minimization of the root mean square deviation $d$ while including translation (t), rotation (R), permutation (P), and inversion (inv) of the atomic coordinates (FASTOVERLAP), followed by specific distinguishing feature identification that helps to define interpretable labeling of the final clusters.}
    \label{fig:umap_hdbscan_algo}
\end{figure*}

\subsection*{Alignment and Permutation}

Structure alignment and permutation is a central component of the procedure we have assembled below. To determine if two complex atomic arrangements sampled from an MD trajectory are similar is easily complicated by the strong likelihood that their centers of mass are displaced or translated from one another and that they may have quite different orientations with respect to a given set of Cartesian axes. In structural biology, where large, complex macromolecules may have quite low symmetry, determining an optimal translation and rotation to best overlap two structures can be accomplished by minimizing the root mean square deviation (RMSD) of the overall atomic structure - a Euclidean metric in the multi-atomic three-dimensional coordinates~\cite{Kufareva2012}:

\begin{eqnarray*}
  \mathrm{RMSD}(\mathbf{X},\mathbf{Y})=\sqrt{\frac{1}{N}\sum_{i=1}^N\sum_{\alpha=x,y,z}|X_{\alpha}^{(i)}-Y_{\alpha}^{(i)}|^2} \ ,
\end{eqnarray*}

where $\mathbf{X}$ and $\mathbf{Y}$ are $N$-dimensional sets of atomic position vectors, assumed to have the same index ordering with respect to $i$. Optimal alignment is achieved by finding the parameters that define a translation vector and rotation matrix that minimize the RMSD.

Surprisingly, aligning much smaller clusters can be more challenging due to the strong possibility that they exhibit somewhat random ordering of the atomic coordinates that define the cluster due to their assembly being driven by non-bonding interactions -- Coulombic attraction and/or van der Waals forces, most likely. This possible reordering of the atomic species can discontinuously alter the RMSD and prevent gradient-based minimization algorithms from optimally aligning pairs of structures. Multiple approaches have been developed in the literature to overcome this issue. Reducing atomic structural information into orientation-independent features (bond lengths, angles, dihedrals, Coulomb matrices~\cite{Rupp2012}, bag of bonds~\cite{Hansen2015}, etc.) can help, but still suffers from the possible random ordering of individual elements~\cite{Elton2018}. Binning or histogramming the atomic data as a density field~\cite{Ceriotti2018} or within some other reduced-dimensional feature vector~\cite{Reinhart2021} can remove this issue but may require large fields to store the information or may be dependent on the grid-spacing of the density/histogram representation. 

Due to the factorial scaling of permutations with respect to the number of identical atoms in the local atomic arrangements, a brute force approach to enumerating and comparing all possibilities may become inefficient or at worst intractable. The so-called Hungarian method~\cite{Kuhn1955}, and approaches derived from it such as the shortest augmenting path~\cite{Jonker1987, Carpaneto1988} provide a solution to this combinatorial optimization problem in polynomial time. 
We make use of FASTOVERLAP's branch and bound alignment algorithm~\cite{Griffiths2017} to align pairs of atomic arrangements by minimizing their RMSD with respect to permutation ($P$) and inversion ($\mathrm{inv}$), in addition to translation (t) and rotation (R) of the 3D structures:
\begin{eqnarray*}
  d_{\mathrm{\min}}(\mathbf{X},\mathbf{Y}) = \min_{t,R,P,\mathrm{inv}} RMSD(\mathbf{X},\mathbf{Y'}) \ ,
\end{eqnarray*}
where $\mathbf{Y'}$ is the translated, rotated, permuted, inverted version of $\mathbf{Y}$.

The final output of optimally aligned structures greatly facilitates a fast appreciation of the salient features of a given cluster of atomic arrangements, so that they might be tagged with physically interpretable labels. 

Computing $d_{\min}$ between every pair of atomic arrangements to define a huge distance matrix (scaling as the square of the number of arrangements) would definitely reveal details of distinct groups within the data set, since this metric defines which arrangements are ``close'' to one another. This distance matrix could be considered as a representation of the underlying distribution or density of the data set. However, this metric scales with the square of the number of local atomic arrangements. Which is still an unwieldy object to compute and to analyze efficiently and automatically for a large dataset.

\subsection*{Clustering algorithm}
In order to classify the local atomic environments into distinct groups, we decide to rely on a procedure that leverages a combination of: the dimensionality reduction technique UMAP~\cite{McInnes2018}, to provide a low dimensional projection of large data sets while maintaining their topological structure; and the hierarchical clustering algorithm HDBSCAN~\cite{McInnes2017}, to organize the low dimensional data into similar groups. Pre-processing the data with UMAP greatly facilitates the clustering procedure, since HDBSCAN understandably struggles to effectively identify dense regions in spaces of higher dimensionality~\cite{McInnes2017}. Our approach is designed to significantly reduce the computational overhead for comparisons via structural alignment and permutation.

A scheme for the analysis procedure is shown in Figure \ref{fig:umap_hdbscan_algo}. After extracting the local atomic arrangements from the trajectory, we separate them into groups, each with a distinct chemical formula (e.g. \ce{B2CaO4}), which we analyze separately.

(1) The first step is to calculate the flattened upper triangular atomic distance matrix for every local atomic arrangement in the trajectory. This results in a row vector containing the pairwise distance between every pair of atoms in the local atomic arrangement. Relying on the flattened distance matrix instead of the raw atomic positions is advantageous, since it is invariant to arbitrary translations or rotations of the local atomic arrangements. 

2) Using UMAP, we generate a low-dimensional projection of the input data -- a reduction from $n=N(N-1)/2$ to 2, where $N$ is the total number of atoms in the local arrangement. Points that are close to each other in the UMAP space have a similar flattened upper triangular distance matrix. This densification in the reduced dimensions of the UMAP space facilitates clustering by HDBSCAN into labelled groups. Arrangements that have been assigned a different label due to significant differences in their distance matrix vectors could actually comprise similar structures that differ only because of an arbitrary permutation of their atomic indexes.

(3) We extract from each HDBSCAN cluster some exemplar local atomic arrangements, viz., particular arrangements that come from high density regions in each cluster. Then, using FASTOVERLAP, we obtain the minimal distance $d_{\min}$ between each pair of exemplars (including permutations) and build a corresponding exemplar distance matrix.

(4) We provide UMAP with this exemplar distance matrix as a feature vector to generate another low-dimensional projection, which we then organize into clusters with HDBSCAN. Since $d_{\min}$ obtained with FASTOVERLAP now accounts for permutation of the atomic indexes, as well as inversion, rotation and translation of the 3D structure, we would expect the total number of groups to reduce significantly at this step, as symmetry equivalent exemplars should merge.

(5) We extract new exemplars from this second UMAP projection space to select an even smaller subset that will constitute an alignment basis set. With the assumption that the sampled trajectory is complete -- i.e., it contains all the expected local atomic arrangements -- the alignment basis set should also be complete. 

(6) We self-consistently align the elements of this basis to one another to more easily compare them upon visualization. Obtaining an optimal global alignment between multiple structures can be challenging. Normally, pairwise alignment algorithms can efficiently align structures two-by-two by minimizing their RMSD. However, minimizing the total RMSD of multiple structures at the same time adds an additional level of complexity. It is important to ensure that the main features of the coordination environment are properly aligned in a similar way. As an example, if we were unfortunate enough to try to align all structures to a single reference that coincidentally has a well-defined symmetry axis, it is possible that structures with equivalent structural features end up aligned differently. The goal of the global optimal alignment is to reduce a cost function consisting of the average RMSD between the exemplar arrangements using our chosen structural alignment procedure. This process is iterative, creating a biased random linkage between the exemplars and sequentially looping through the linkage to align each structure pairwise. At each step, the new alignment is accepted only if the new average RMSD between all exemplars is lower than the previous one. 

(7) We now use the globally aligned exemplar atomic arrangements as reference points: we loop through the entire trajectory and try to align each local atomic arrangement to the exemplars, only keeping that alignment with the smallest $d_{\min}$. This step ensures that all arrangements have a consistent atomic indexing with respect to each other. Only the alignment with the smallest $d_{\min}$ is kept for each local atomic arrangement, but we subsequently compute its RMSD (i.e., without alignment or permutation) with respect to all elements of the alignment basis to define a new feature vector.

(8) At this point, to perform a final clustering iteration, we are left with three possibilities. Since now the atomic indexing is consistent, we can calculate the flattened upper triangular atomic distance matrix once more and provide it to UMAP. Alternatively, since the vector of $d_{\min}$ between each local atomic arrangement and the alignment basis is unique, it can be used as a direct input for dimensionality reduction. Finally, a combination where we provide both the flattened upper triangular atomic distance matrix and  the vector of $d_{\min}$ to the alignment basis set can be used. In all three cases, the low dimensional projection obtained with UMAP is then given to the HDBSCAN algorithm. The final product is distinct clusters of structures that have similar local atomic arrangements and that have been optimally aligned with each other.\\

%% file: Sections/03_computational_details.tex
\section*{Computational Details}

The model system of this study comprises one \ce{Ca^{2+}} cation and two \ce{BH_4^-} anions dissolved in THF next to a graphite interface. THF is an organic solvent, a molecular liquid with the formula \ce{C4H8O}, with the heavy atoms arranged in a pentagonal ring and two hydrogen atoms bound to each carbon. There are 336 THF molecules and two layers of 416 carbon atoms defining the graphite surface. The initial configuration was generated by building a simulation box of size $a=34.08$ \AA{}, $b=31.97$ \AA{}, $c=50.00$ \AA{}, where $c$ is perpendicular to the plane of the graphite surface, with the help of the software PACKMOL~\cite{Martinez2009}.\\

All classical molecular dynamics simulations were ran with a timestep of 1 fs and the OPLSAA force field~\cite{Samba2009}. The choice of \ce{BH_4^-} charges was based on comparisons with Density Functional Theory interaction energies between a \ce{Ca^{2+}} cation and a \ce{BH_4^-} anion at face, vertex and edge sites. NBO charges provided the best agreement. Additionally, charge scaling by a factor of 0.7 was applied in order to correctly reproduce the ab-initio free-energy profile of \ce{Ca^{2+}} in THF as a function of coordination~\cite{Baskin2020, Driscoll2020} using the methods described below. The particle-mesh Ewald (PME) method with a 1.0 nm cut-off distance and  $10^{-5}$ grid spacing in $k$-space were used to treat long-range electrostatic interactions. The cut-off for the Lennard-Jones interactions was 1.0 nm and the spline ranged from 0.9 nm to 1.0 nm.\\

To equilibrate the initial simulation box, we performed multiple steps. First, the structure was relaxed using steepest descent followed by conjugate gradient minimization with a tolerance of $10^{-4}$ kcal mol$^{-1}$ for the energies and $10^{-4}$ kcal mol$^{-1}$ \AA$^{-1}$ for the forces. Then the system was heated to 298 K during 10,000 steps (10 ps) of NVT simulation. NPT dynamics with an isotropic pressure of 1 atm were then run for 1 ns using a Nose/Hoover thermostat~\cite{Nose1984} and a Nose/Hoover barostat~\cite{Hoover1985}. The temperature coupling constant was 0.1~ps and the pressure piston constant was 2.0~ps. The equilibrated lattice parameters were obtained by averaging the box size over the last 0.5 ns of NPT trajectory. Finally, we performed an additional thermalization at 298 K with the newly calculated lattice parameters under NVT dynamics for 5.6~ns.\\

The evaluation of the free energy profile as a function of the two collective variables (CV) -- the coordination number (CN) between calcium and the oxygen of the THF molecules and the distance from the outermost graphite layer (dZ) -- was performed using the COLVARS module~\cite{Fiorin2013} as implemented in LAMMPS~\cite{LAMMPS}. The initial configuration consisted of equilibrated simulation cells of solvent molecules with ions as described above, and the same MD parameters were used as in the final equilibration step. The cut-off radius for the coordination number was 3.6 \AA{}. To prevent the calcium ion from moving too far away from the graphite, a harmonic potential wall was set at dZ = 24~\AA{}. The free energy surface is stored on a discrete grid with a spacing of 0.05 (unitless and \AA{} for the two CVs, respectively). Biasing was applied every 200 time steps by adding Gaussian hills of weight 0.02 kcal/mol and width twice the grid spacing for the two CVs. To accelerate the sampling we used multiple (14) walker metadynamics~\cite{Raiteri2006} communicating every 25,000 timesteps. The trajectory of each replica comprises 300~ns, totalling approximately 4.2~$\mu$s.\\ 

Umbrella sampling was performed using the same MD parameters as during the equilibration and metadynamics run by adding a single harmonic umbrella potential at a well defined point on the CV surface. At each point, a trajectory length of 70~ns was collected. The trajectory was saved every 5~ps for further analysis.\\

Dimensionality reduction with UMAP~\cite{McInnes2018} was performed first by removing the mean and scaling to unit variance the data~\cite{scikit-learn} and subsequently embedding the scaled data in a two-dimensional space. The size of the local neighborhood was constrained to 15 neighbors with a minimum distance between points of 0.0 and the distance was computed with the Euclidean metric. Hierarchical clustering with HDBSCAN~\cite{McInnes2017} was performed on the low-dimensional projection with standard parameters and allowing for the possibility of there being only a single cluster. In some cases, parameters were tweaked iteratively depending on visual feedback to obtain reasonable clustering. Namely, varying the number of neighbours in the UMAP projection between 10 and 30 and increasing the value of the smallest size grouping for the HDBSCAN clusters.

Structural alignment was performed with the branch and bound algorithm implemented in the FASTOVERLAP package~\cite{Griffiths2017}.

Our clustering algorithm was implemented in Python, leverages available open-source data science modules (NumPy~\cite{numpy}, sklearn~\cite{scikit-learn}) and integrates with the Atomic Simulation Environment (ASE)~\cite{Larsen2017}. The plots were generated with the help of Matplotlib~\cite{matplotlib} and seaborn~\cite{seaborn}.

%% file: Sections/04_results.tex
\section*{Results}

\begin{figure}[ht]
    \centering
    \subfloat[Free energy surface\label{subfig:metadyn_coordO_dZ}]{
        \includegraphics[width=\columnwidth]{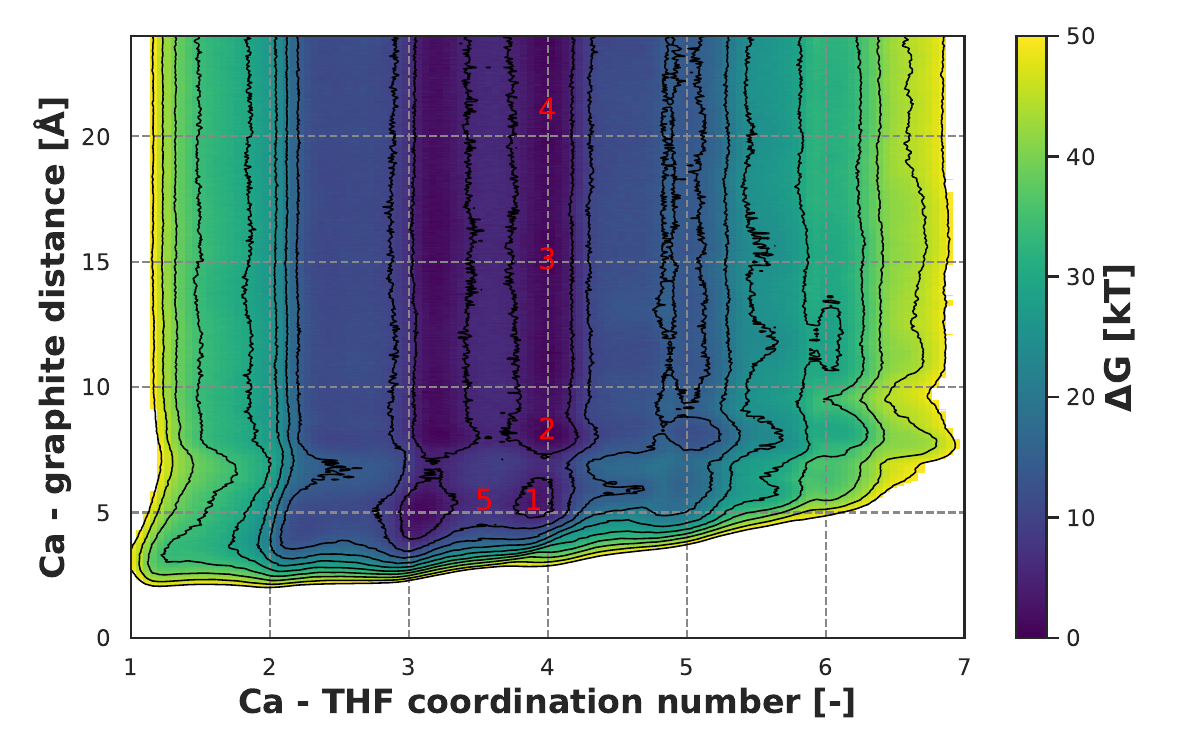}
        }
        
    \subfloat[RDF plots\label{subfig:RDF_plots}]{
        \centerline{\includegraphics[width=\columnwidth]{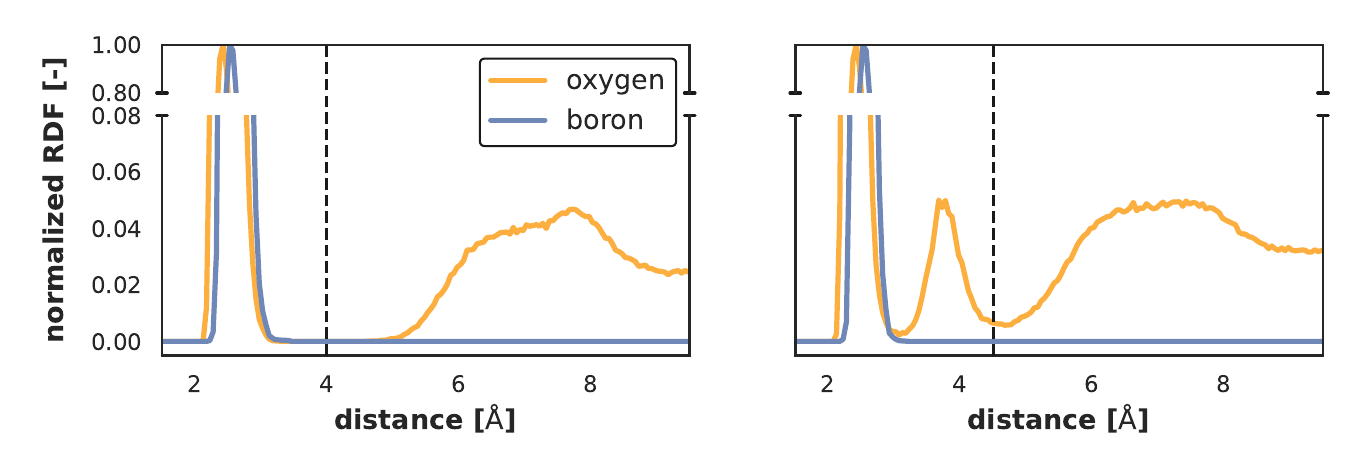}
        }}
        \caption{(a) Two-dimensional free energy surface for the solvation of \ce{Ca^{2+}} in THF at a graphite interface as a function of the collective variables: x) coordination number between the Ca and the O of the THF atoms, y) distance between the Ca and the outermost graphite layer. The numbers 1-5 correspond to the region sampled with umbrella sampling and used in the results section. (b) Normalized O and B RDF plots around Ca for trajectories 1-4 (left) and trajectory 5 (right). The vertical dashed line corresponds to the cutoff radii for the analysis discussed later.}
    \label{fig:metadyn_coordO_dZ_and_RDF_plot}
\end{figure}

\subsection*{Molecular Dynamics Sampling}

We provide here some representative examples of applying the clustering procedure to an MD trajectory. Our chosen application is to a cutting-edge electrolyte: calcium borohydride [Ca(BH$_4$)$_2$] dissolved in tetrahydrofuran (THF). We first run a metadynamics simulation to calculate the two-dimensional free energy landscape as a function of two collective variables: the Ca-O coordination number and the distance of the calcium ion from the graphite interface (see Computational Details). The free energy surface shown in Figure~\ref{subfig:metadyn_coordO_dZ} exhibits multiple distinct minima, with the most prominent for Ca-O coordination numbers of approximately 3 and 4 that extend as valleys at higher distances from the graphite. Energy barriers on the order of 5-10~k$_B$T at 300~K separate the minima in these valleys. The lowest point on the energy surface is point 2. Points 1, 3, 4 and 5 have a free energy of 3.1~k$_B$T, 0.7~k$_B$T, 0.9~k$_B$T and 6.9~k$_B$T higher respectively. We are interested in identifying similarities and differences in the calcium coordination environment at different locations across the free energy landscape. With our initial metadynamics map of the free energy landscape, we next employ umbrella sampling, a constrained MD protocol biased by an additional harmonic potential, centered at particular coordinates on the space spanned by our collective variables. The harmonic potential ensures a local sampling of the MD trajectory close to the free energy surface and allows us to extract quantitative information regarding the relative local ionic populations (see Computational Details).\\

We choose five distinct regions where we run umbrella sampling. Locations 1, 2, 3 and 4 are at different distances from the graphite interface, but all in the free energy valley with Ca-O coordination number of approximately 4. Location 5 is at the transition point between two minima with coordination numbers 3 and 4, respectively, that are the closest to the graphite. Depending on the location of the harmonic constraint with respect to these collective variables, distinct chemical arrangements around \ce{Ca^{2+}} can be observed. We are particularly interested in the first coordination shell around the ion. Being positively charged, Ca ions attract the oxygen atom in THF and the \ce{BH_4^-} anions. In each case, the local atomic arrangement can be assigned a chemical formula that indicates its composition in terms of the number of each atomic species.

\begin{figure}[ht]
    \centering
    \subfloat[UMAP projection\label{subfig:B2CaO4_UMAP}]{
    \centerline{
            \includegraphics[width=\columnwidth]{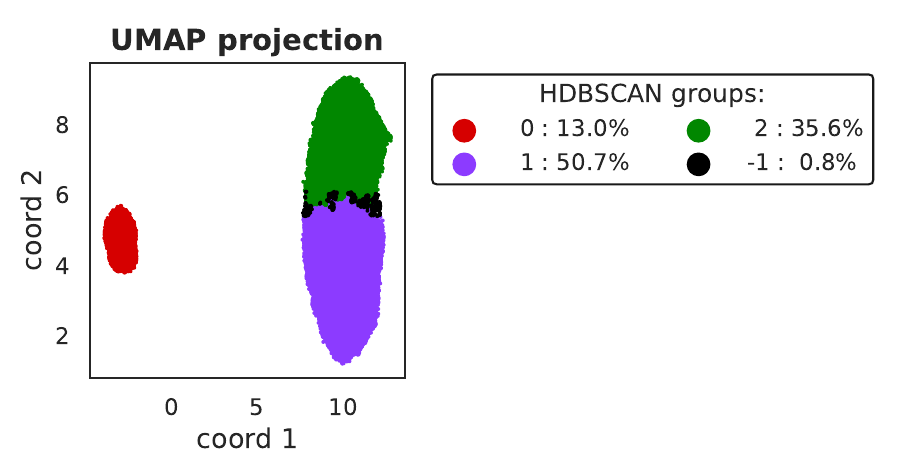}
            }}
        
    \subfloat[Top view\label{subfig:B2CaO4_MS_top}]{
        \centerline{\includegraphics[width=\columnwidth]{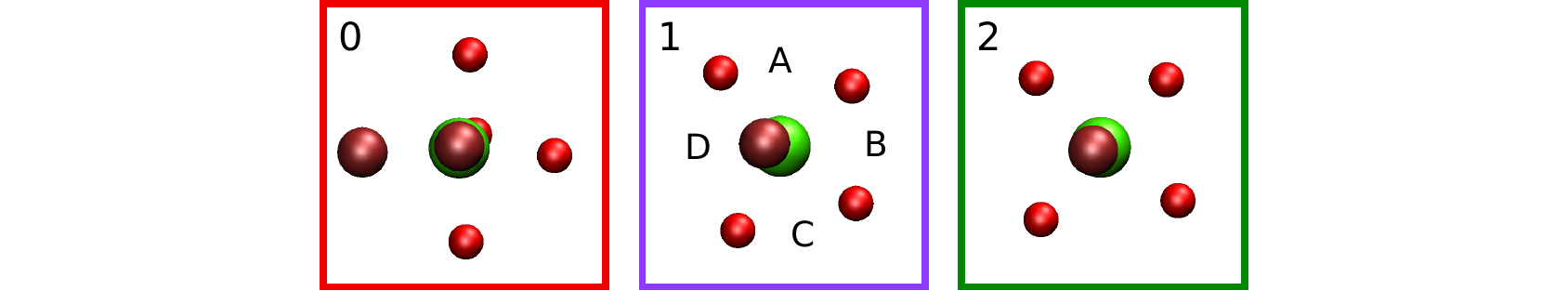}
        }}
        
    \subfloat[Side view\label{subfig:B2CaO4_MS_side}]{
        \centerline{\includegraphics[width=\columnwidth]{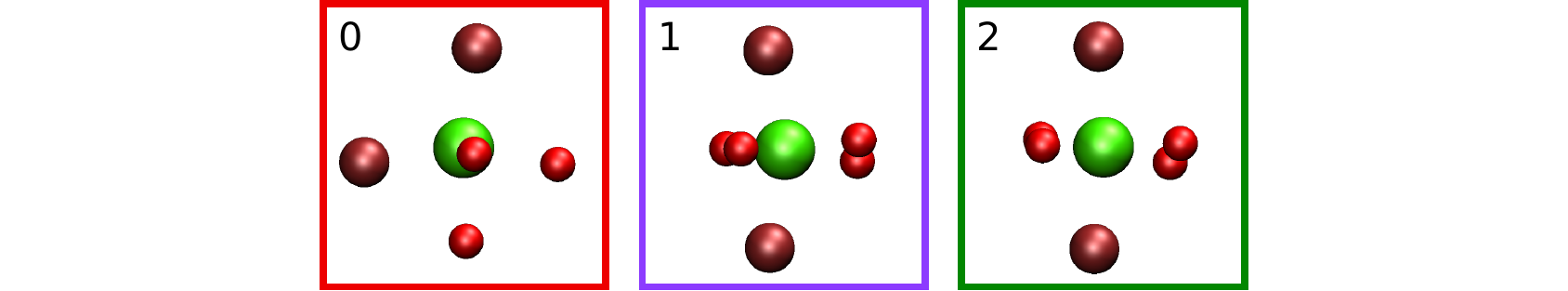}
        }}
    \caption{(a) UMAP projection colored according to HDBSCAN labelling of \ce{B2CaO4} local atomic arrangements of trajectories 1-4. Points classified as noise are labelled -1 (black). (b) Top and (c) side view of the averaged local atomic arrangements for each HDBSCAN group.}
    \label{fig:B2CaO4_UMAP_and_mean_structures}
\end{figure}

\subsection*{Local Atomic Arrangements}

First, we analyze minima 1-4 along the valley with a coordination number of 4. We are interested in knowing the difference in local atomic arrangement as a function of the distance from the graphite interface. Initially, we limit our analysis to only three atomic species (Ca, O, and B), ignoring details of the solvent molecule or anion orientation. To select an appropriate cutoff radius and extract the local atomic arrangements from the trajectories, we use the radial distribution functions (left RDF plot in Figure~\ref{subfig:RDF_plots}). The RDFs of O and B with respect to Ca show two narrow peaks at 2.35~\AA{} and 2.45~\AA{}, respectively, indicating the extent of the first solvation sphere around the cation. Beyond these first peaks, the RDF is almost flat until approximately 5~\AA{}, where the oxygen atoms of the second solvation sphere appear. By choosing a cutoff radius of 4~\AA{} for both B and O, we can effectively isolate the local atomic environment around \ce{Ca^{2+}} and ensure we only include the first solvation shell. The predominant total coordination number (including boron and oxygen) for calcium in this environment is 6, but other coordination numbers could occur. For example, at the neighboring minima with a coordination number of 3 the maximum total coordination number allowed is 5 (since there are only two available borohydride ions in the entire simulation box). In the umbrella sampling, we constrain 4 oxygen atoms to be coordinated to the calcium and since the two available borohydride anions are strongly attracted by the divalent cation, we observe only one chemical formula at the four sampled points, namely \ce{B2CaO4}. After extracting the local atomic arrangements, each sampling point provides 12,000 molecular snapshots, totalling 48,000 atomic arrangements. We then proceed to employ the clustering algorithm to obtain the main coordination environments and compare their occurrence as a function of their distance from the graphite interface without previous assumptions on their arrangements.

\begin{figure}[ht]
    \centering
    \includegraphics[width=0.8\columnwidth]{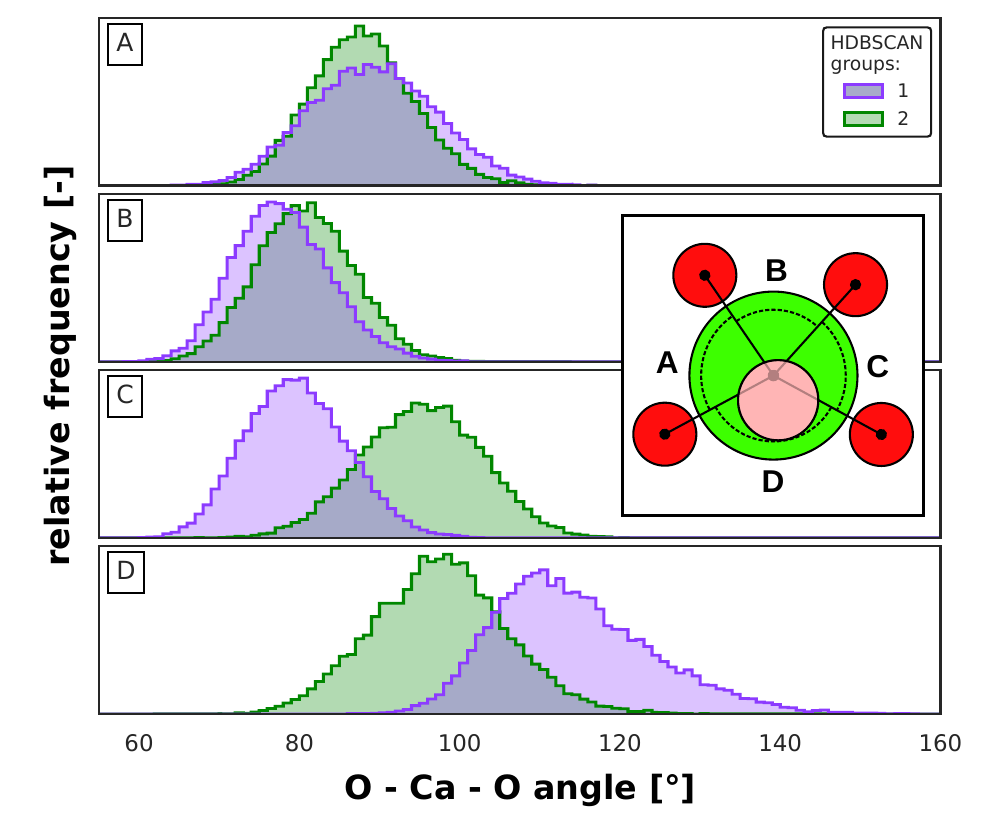}
    \caption{Distribution of angles between neighbouring O atoms with center on the Ca for the groups 1 and 2. The detail within the figure shows the ordering of the angles A, B, C and D.}
    \label{fig:B2CaO4_O-Ca-O_angles}
\end{figure}

\begin{figure*}[ht]
    \centering
    \includegraphics[width=0.8\linewidth]{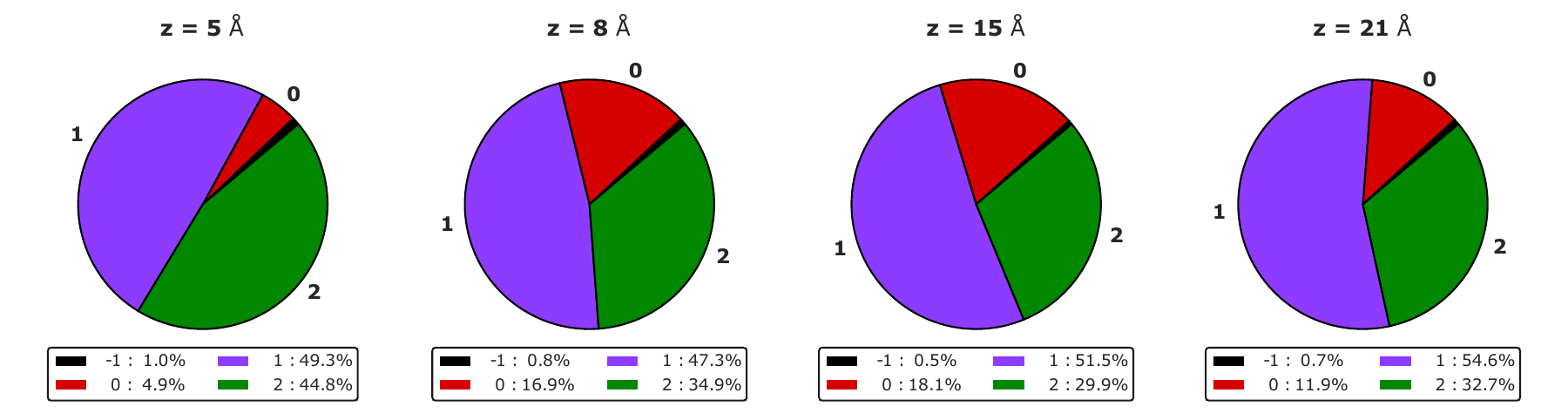}
    \caption{Distribution of species of \ce{B2CaO4} as a function of distance from the interface (z).}
    \label{fig:B2CaO4_pies}
\end{figure*}

\subsection*{Clustering Analysis}

The clustering algorithm identifies 3 main local atomic arrangements across the four distances sampled. The final projection of the UMAP space on 2 dimensions with the relative clusters labelled with HDBSCAN is shown in Figure~\ref{subfig:B2CaO4_UMAP}.

The projected coordinates are mainly divided into two regions in the UMAP space. On the right, a big cluster of points is divided into two groups (1 and 2). On the left, there is the isolated group 0. The black points labeled -1 which HDBSCAN did not classify as belonging to any group are considered noise, although they are most likely intermediate arrangements between groups 1 and 2. The most occurring atomic arrangement, group 1, accounts for roughly 50.7\% of the 48,000 arrangements analyzed. The second most occurring, group 2, accounts for 35.6\% of the arrangements. Group 0 consists of 13.0\% of the arrangements. Only 0.8\% of the data was labelled as noise. The reference atomic structures of each labelled group are shown in Figures~\ref{subfig:B2CaO4_MS_top} and \ref{subfig:B2CaO4_MS_side}.\\

To try to understand why the local atomic arrangements were divided into different groups, we look at the angles between the oxygen atoms and the boron around the central calcium atom. The biggest difference comes from the bent or linear arrangement of the two B atoms. In group 0, the B atoms are adjacent, as proposed in \cite{Hahn2020}, while in groups 1 and 2 they are located axially on opposite sides of the Ca ion. Further differences of configuration between groups 1 and 2 arise from differences in the O-Ca-O angles.

Figure~\ref{fig:B2CaO4_O-Ca-O_angles} shows the four O-Ca-O angles of groups 1 and 2. It becomes evident that group 1 consists of local atomic arrangements where one of the O-Ca-O angles is much larger than the average. Specifically, the mean angle D is $114.3^{\circ} \pm 9.8^{\circ}$ and $97.5^{\circ} \pm 8.3^{\circ}$, in groups 1 and 2, respectively. The fact that there seem to be normal distribution of angles with distinct means for the different groups indicates that those features play a role in defining the final UMAP projection of the local atomic arrangements. Specifically, since we only provide UMAP with the distances between atoms in a local atomic arrangement, the O-O distances responsible for the O-Ca-O angle are distinct between different groups.

We can see from Figure~\ref{fig:B2CaO4_pies} that depending on the distance of the calcium ion from the graphite, the relative population of the three coordination environments changes. Closest to the interface, only 4.9\% of the coordination structures belong to group 0 (bent configuration). This number increases to 16.9\% in the second layer, reaches the highest value of 18.1\% at 15~\AA{} from the graphite and finally decreases to 11.9\% in the bulk. Interestingly, the ratio between local arrangements 1 and 2 is highest at the interface, where their prevalence is almost equal. Most likely, local atomic arrangement 2 is favoured by the presence of the graphite. At all distances sampled, approximately 50\% of the structures belongs to group 1.\\

The clustering procedure involving UMAP and HDBSCAN was able to separate not only atomic arrangements that are clearly different (adjacent vs opposite boron arrangement), but also atomic arrangements with finer differences of angles between the coordinating atoms (O-Ca-O angle). However, we noticed by repeating the analysis that it is not always possible to distinguish the atomic arrangements with small O-Ca-O angle differences (groups 1 and 2). The main reason behind this is the randomness embedded in the dimensionality reduction and clustering pipeline. UMAP is a stochastic algorithm and therefore some variance in the results is to be expected, although a consistent random seed can be set to guarantee reproducibility for a given data set. There is also some stochasticity in our iterative global alignment procedure, based on its choice of random linkage. In general we observed the low dimensionality projection to be quite stable. However, mostly when working with smaller feature vectors (considering less atoms) and fine structural differences in atomic arrangement, a different final clustering could be obtained. Despite this, the differentiation of the main feature -- the B-Ca-B angle -- is always retrieved, which is a good indicator that the clustering algorithm functions when the local atomic arrangements are clearly distinct.

\begin{figure}[ht]
    \centering
    \subfloat[UMAP projection\label{subfig:C16B2CaO4_UMAP}]{
        \centerline{
        \includegraphics[width=\columnwidth]{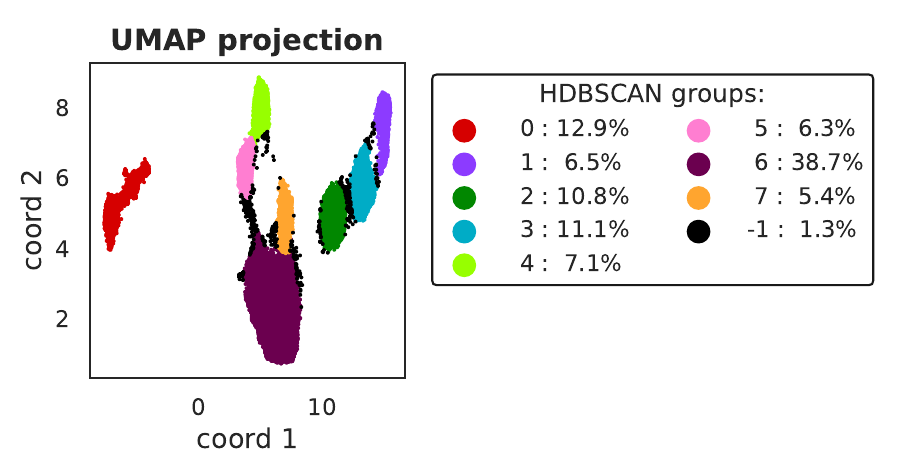}
        }}
        
    \subfloat[Top view\label{subfig:C16B2CaO4_MS_top}]{
        \centerline{
        \includegraphics[width=\columnwidth]{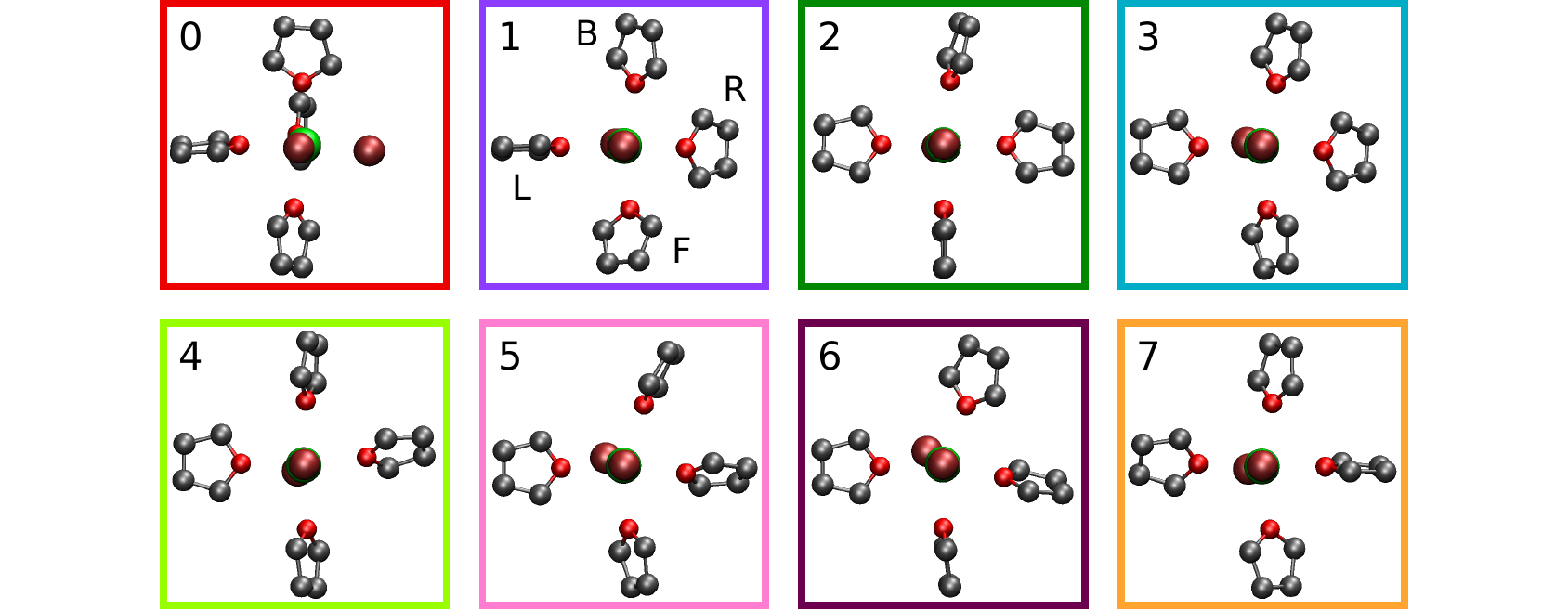}
        }}
        
    \subfloat[Side view\label{subfig:C16B2CaO4_MS_side}]{
        \centerline{
        \includegraphics[width=\columnwidth]{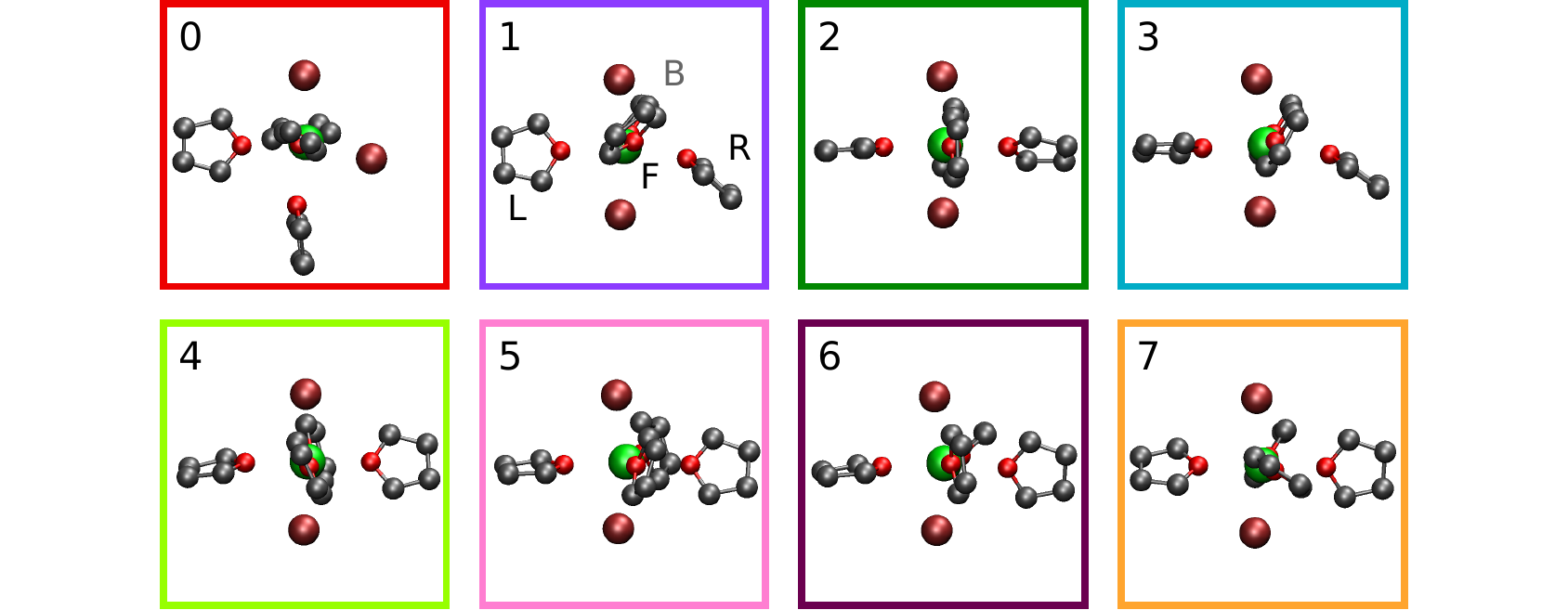}
        }}

    \caption{(a) UMAP projection colored according to HDBSCAN labelling of \ce{B2CaO4} local atomic arrangements that include THF backbones of trajectories 1-4. Points classified as noise are labelled -1 (black). (b) Top and (c) side view of the averaged local atomic arrangements for each HDBSCAN group. The four THF molecules in the configurations with the boron axially arranged (groups 1-7) are B: back, F: front, L: left, R: right.}
    \label{fig:C16B2CaO4_UMAP_and_mean_structures}
\end{figure}

\begin{figure}[ht]
    \centering
    \includegraphics[width=\columnwidth]{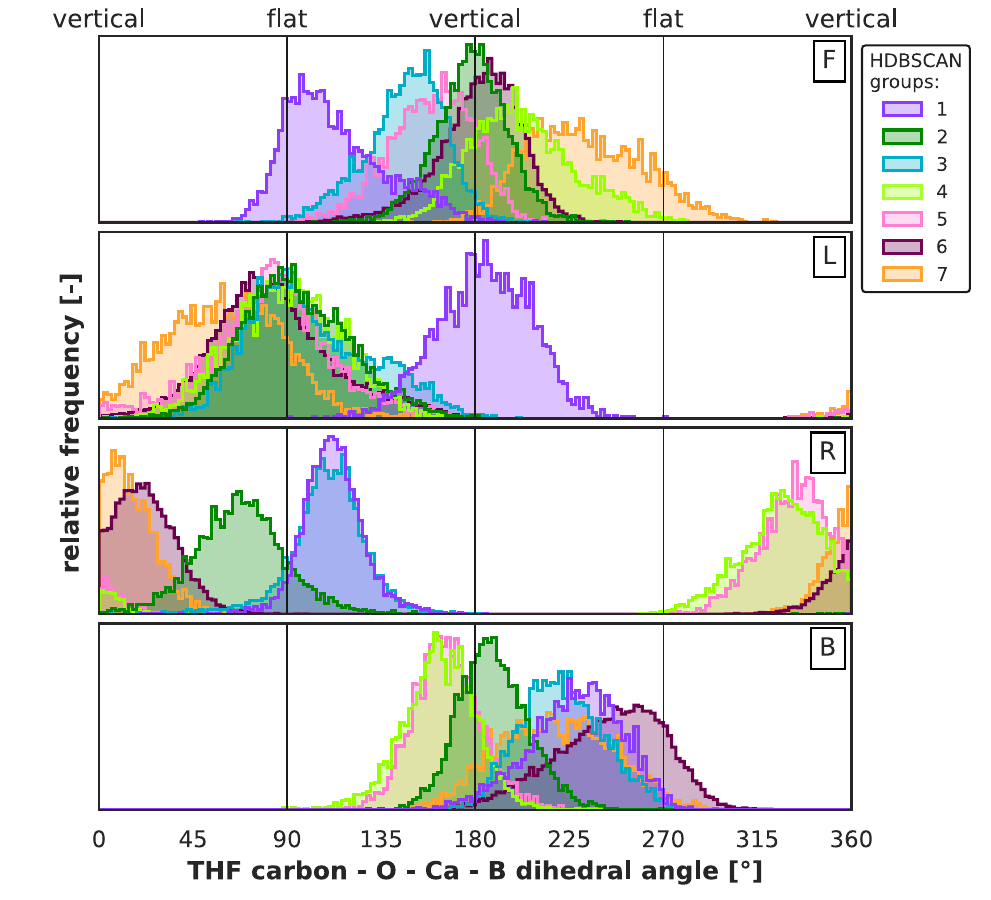}
    \caption{Distribution of the dihedrals between the C atoms of the THF molecules and the axis spawned by the two B atoms in the case they are axially arranged (groups 1-7 in Figure~\ref{fig:C16B2CaO4_UMAP_and_mean_structures}). F: front, L: left, R: right, B: back THF molecule.}
    \label{fig:B2CaO4_THF_BB_dihedrals}
\end{figure}

\begin{figure}[ht]
    \centering
    \includegraphics[width=\columnwidth]{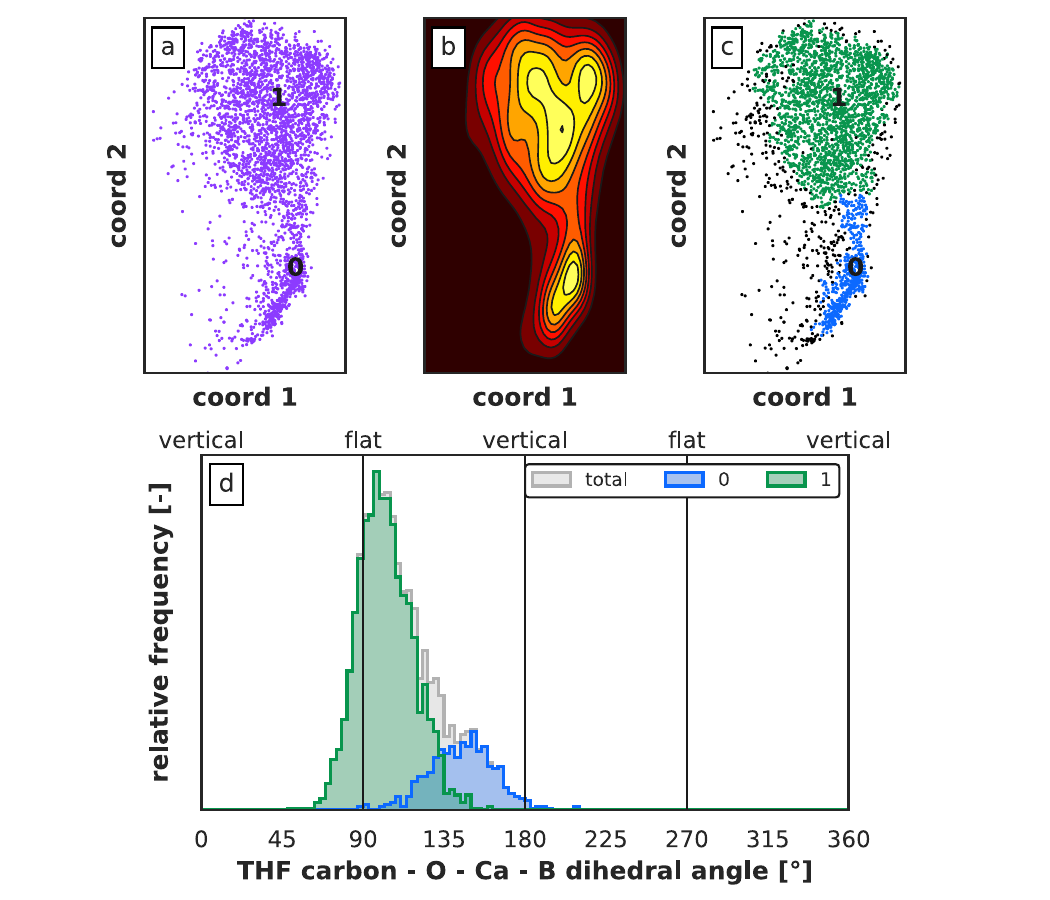}
    \caption{(a) UMAP projection of group 1 from Figure~\ref{fig:C16B2CaO4_UMAP_and_mean_structures}. The labels 0 and 1 correspond to the peaks in density of the heatmap in (b). (c) Contribution of the two regions to the first dihedral angle of group 1 (F) in Figure~\ref{fig:B2CaO4_THF_BB_dihedrals}. (d) Dihedral of the F THF molecule with contribution from the two high density regions.}
    \label{fig:C16B2CaO4_g1_dihedrals_detail}
\end{figure}

\begin{figure*}[ht]
    \subfloat[Orientation with respect to graphite\label{subfig:g1_g3_interface}]{
    \centering
    \includegraphics[width=0.13\linewidth]{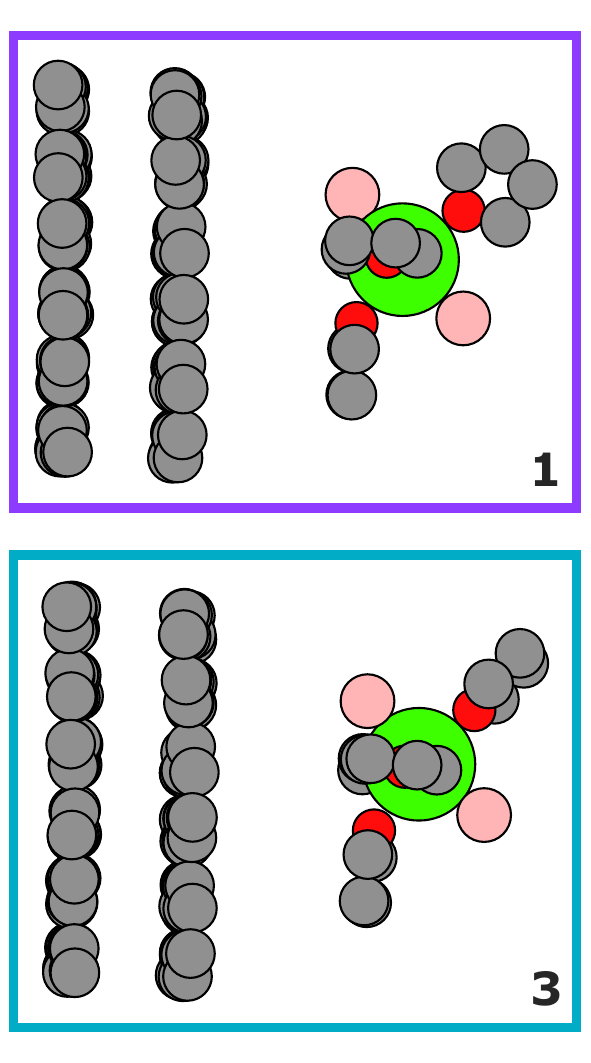
    }}\quad
    \subfloat[Distribution of species\label{subfig:C16B2CaO4_pies}]{
    \centering
    \includegraphics[width=0.8\linewidth]{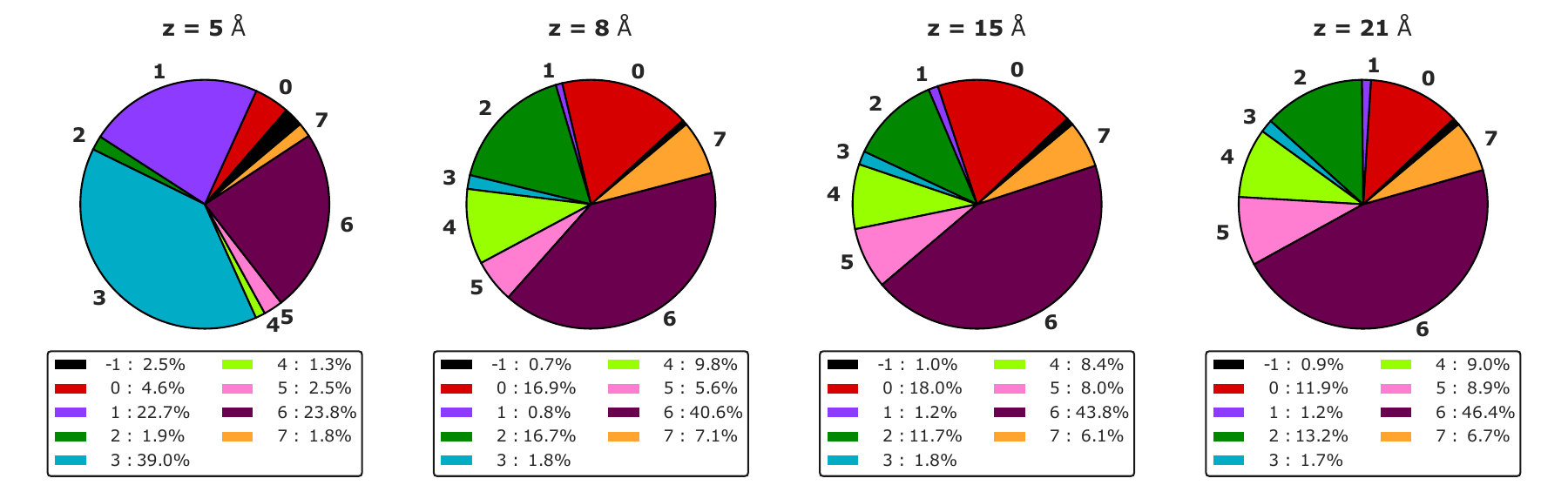
    }}
    \label{fig:g1_g3_interface_and_C16B2CaO4_pies}
        \caption{(a) Orientation of local atomic arrangements of isomers 1 and 3 with respect to the graphite layer. (b) Distribution of species of \ce{B2CaO4} with THF carbons as a function of the distance to the interface (z). Note that isomers 1 and 3 appear almost uniquely closest to the surface.}
\end{figure*}

\subsection*{Increasing Complexity}

In order to obtain a more detailed picture of the chemistry around calcium in the MD trajectory, we decided to increase the size of the local atomic arrangements by including the carbon atoms of the coordinated THF (while still ignoring the hydrogen atoms of the borohydride anion and THF). Here we do not explicitly use a cutoff radius for the carbon atoms, instead we use the molecular bonding information provided to LAMMPS to reconstruct the THF molecules with an oxygen atom within the cut-off radius of 4~\AA{} from the specified calcium ion. After re-running the clustering procedure, we can observe even richer information on the local coordination environment (Figure~\ref{subfig:C16B2CaO4_UMAP}).\\

This time we identify 8 main groups using the clustering procedure.  One group (group 0) is distant from the other ones on the UMAP projection space. The local atomic arrangement of group 0 exhibits two adjacent B atoms, while groups 1-7 all have the B atoms axially arranged on opposite sides of the Ca cation. The total occurrence of group 0 is 12.9\%, while in the previous analysis, where we still were excluding the carbon atoms, it was 13.0\%. This indicates that both with and without the THF carbon atoms the UMAP projection effectively isolates this structural characteristic -- a bent B-Ca-B configuration. The remaining 87.1\% of the atomic arrangements belong to cases where the boron atoms are on opposite sides of the calcium ion. In this configuration, the THF molecules are equatorially arranged on a plane around the cation. These local atomic arrangements are divided by our algorithm into 7 separated groups in the UMAP space. The main differences between those groups lie in the orientation of the THF molecules around the calcium. 

The mean structure of the local atomic arrangements belonging to each HDBSCAN group is shown in Figures~\ref{subfig:C16B2CaO4_MS_top} and \ref{subfig:C16B2CaO4_MS_side}. Apart from the clear difference between group 0 and the rest, we can see how the clustering procedure succeeds in revealing different arrangements of the THF molecules around the calcium ion. From the visualization of the mean structures, we learn how the COCaB dihedral angle formed by one carbon atom adjacent to an oxygen atom in the same THF molecule and the the Ca-B axis seems to be a relevant metric. The distribution of these COCaB dihedral angles is shown in Figure~\ref{fig:B2CaO4_THF_BB_dihedrals}. The presence of approximately normal COCaB dihedral distributions (in that they are unimodal) indicates that indeed there are well defined mean COCaB dihedral angles for each group of local atomic arrangements. Furthermore, this validates the choice of averaging the atomic positions to obtain a mean structure, given that the mean structures appear physically reasonable. This is a strong indication that the clustering process has revealed and separated normally distributed data with distinct means.\\

Since groups 1-7 all have B atoms arranged axially on opposite sides of the calcium ion, then a COCaB dihedral angle of 0$^{\circ}$ or 180$^{\circ}$ corresponds to the THF molecule with its pentagonal plane aligned with this axis, which we label as vertical. A COCaB dihedral angle of 90$^{\circ}$ or 270$^{\circ}$ we will label as a flat alignment of the THF molecular plane. Angles in between these limiting cases can be referred to as "tilted". Finally, we notice that in most cases the THF molecules are arranged radially with respect to the Ca ion at the center of the atomic arrangement, with the THF dipole vector antiparallel to a radial vector. In those rarer cases where this is not so (dihedral R in 1 and 3), we can label the alignment as non-radial. 

The first thing to be noticed is that no dihedral distribution is similar for the four THF molecules at the same time, which indicates that each group found by the clustering procedure is unique. All configurations with the axial boron have a mix of flat, tilted and vertical coordinating THF molecules. Since the local atomic arrangements have been aligned with each other, the key differences are easily recognizable. Groups 1 and 3, both show a flat, non-radial right THF molecule and only differ from the orientation of the left THF which is vertical and flat respectively. Groups 4 and 5 differ mainly from the orientation of the front and back THF, which are tilted towards the flat THF on the left in one case and towards the vertical THF on the right otherwise. Group 2 has alternating flat and vertical THF molecules, while groups 6 and 7 have a mixture of tilted and flat THFs. Interestingly, we can note how there is no configuration with all four flat THF molecules (or even two adjacent ones). Indeed, ultimately it is sterical hindering that dictates which configurations are possible.

In some cases we notice how some of the COCaB dihedral angle distributions have tails. For example, the dihedral angle of the front THF molecule (F) of group 1 extends from its peak near the flat orientation towards the vertical orientation. To investigate this particular feature, we isolate the local atomic arrangements of this group to perform a more detailed analysis. In Figure~\ref{fig:C16B2CaO4_g1_dihedrals_detail}, we can see indeed how there is a finer substructure within the UMAP projection space that was not previously captured when we ran HDBSCAN on the entire low-dimensional space. Instead, if we run HDBSCAN uniquely on a portion of the UMAP projection -- namely only the data belonging to group 1 -- we are able to separate the former group into two subgroups. Indeed, there are two regions of high data density in the projected space that can clearly be seen by eye in the associated heatmap (Figure~\ref{fig:C16B2CaO4_g1_dihedrals_detail}b). By partitioning the original cluster into these two regions, and plotting the F dihedral again, we can see how this time the corresponding distribution is separated into two distinct groups that appear unimodal.\\

As we have shown for the "coarser" clustering on the \ce{B2CaO4}, the relative population of the groups along the low energy valley of coordination number 4 varies depending on the distance to the graphite. Figure~\ref{subfig:C16B2CaO4_pies} shows the percentage population of each group at the four umbrella sampled points. Again, we can see how the electrolyte/graphite interface shows characteristic atomic arrangements that are more dominant in its immediate vicinity than in regions only 8~\AA{} away or greater. Specifically, the indicated groups 1 and 3 are almost exclusively found in the 5~\AA{} data. When comparing this finding with the mean structures in Figures~\ref{subfig:C16B2CaO4_MS_top} and ~\ref{subfig:C16B2CaO4_MS_side}, we notice how these two local atomic arrangements have one THF molecule (R) inclined so that it is no longer radially oriented with respect to the Ca ion. Figure~\ref{subfig:g1_g3_interface} shows representative local atomic arrangements for groups 1 (top) and 3 (bottom) and their relative orientation with respect to the graphite surface. The local atomic arrangement is rotated such that the non-radial THF molecule lies parallel to the surface plane. The presence of the interface indeed presents a hard physical barrier to the local atomic arrangements. Clearly, there is some flexibility of the coordination shell around calcium that allows the THF molecules and the borohydride to adjust their arrangement to better accommodate the interface. However, the distortion of the THF coordination sphere costs energy to the system. These non-radial arrangements may be the origin of the overall increase in free energy of 3.1~kT at point 1 (5~\AA{} from the graphite surface) with respect to the other sampled points (2-4) within the same valley in Figure~\ref{subfig:metadyn_coordO_dZ}.

\begin{figure}[ht]
    \centering
    \subfloat[UMAP projection\label{subfig:TS_C12B2CaO3_UMAP}]{
    \centerline{
        \includegraphics[width=\columnwidth]{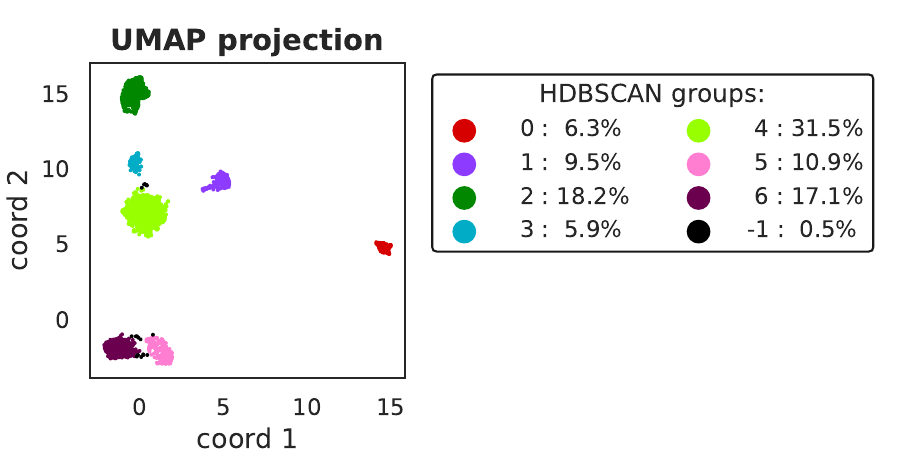}
        }}
        
    \subfloat[Top view\label{subfig:TS_C12B2CaO3_MS_top}]{
        \centerline{
        \includegraphics[width=\columnwidth]{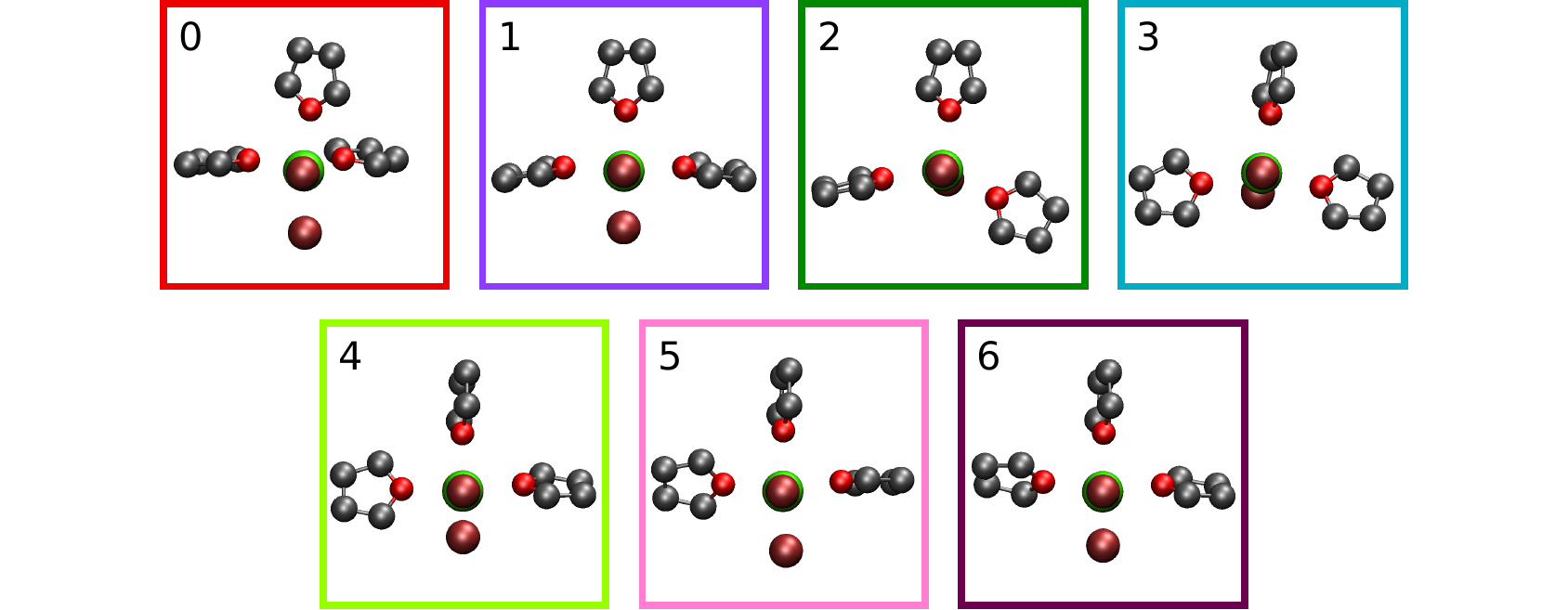}
        }}
        
    \subfloat[Side view\label{subfig:TS_C12B2CaO3_MS_side}]{
        \centerline{
        \includegraphics[width=\columnwidth]{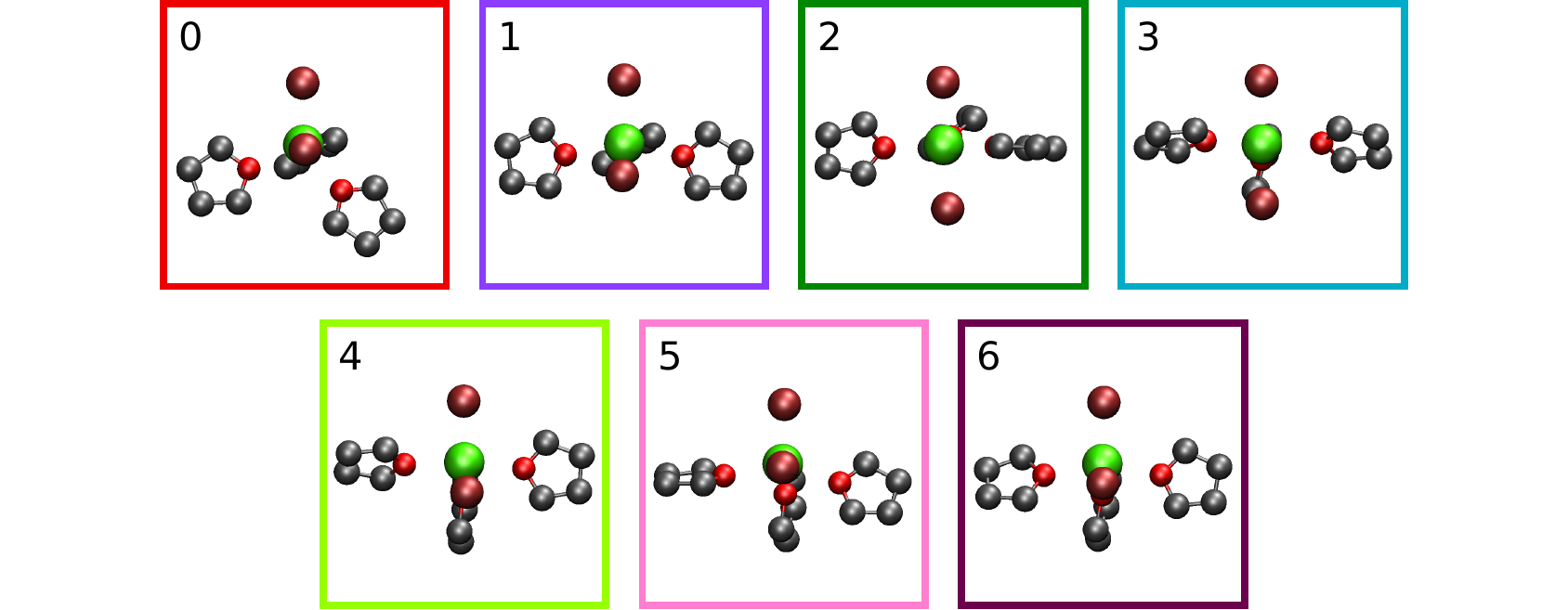}
        }}

    \caption{(a) UMAP projection colored according to HDBSCAN labelling of \ce{C12B2CaO3} local atomic arrangements of trajectory 5. Points classified as noise are labelled -1 (black). (b) Top and (c) side view of the averaged local atomic arrangements with chemical formula \ce{C12B2CaO3} for each HDBSCAN group.}
    \label{fig:TS_C12B2CaO3_UMAP_and_mean_structures}
\end{figure}

\begin{figure}[ht]
    \centering
    \subfloat[UMAP projection\label{subfig:TS_C16B2CaO4_UMAP}]{
        \centerline{
        \includegraphics[width=\columnwidth]{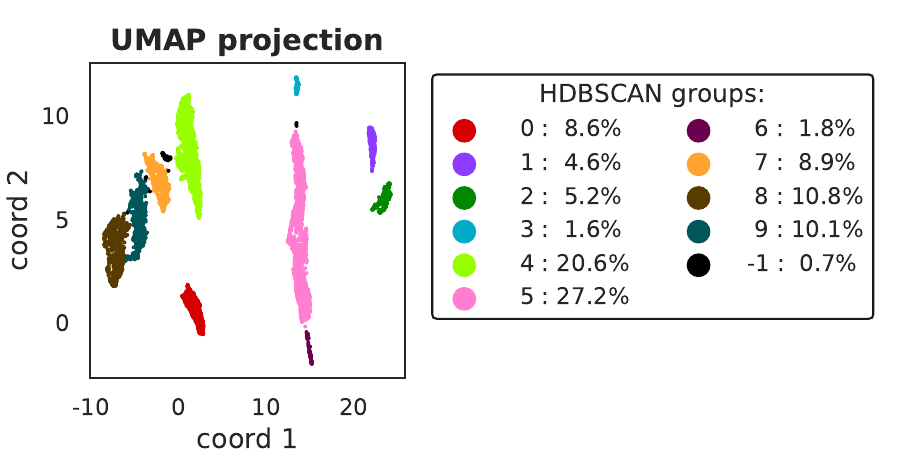}
        }}
        
    \subfloat[Top view\label{subfig:TS_C16B2CaO4_MS_top}]{
        \centerline{
        \includegraphics[width=\columnwidth]{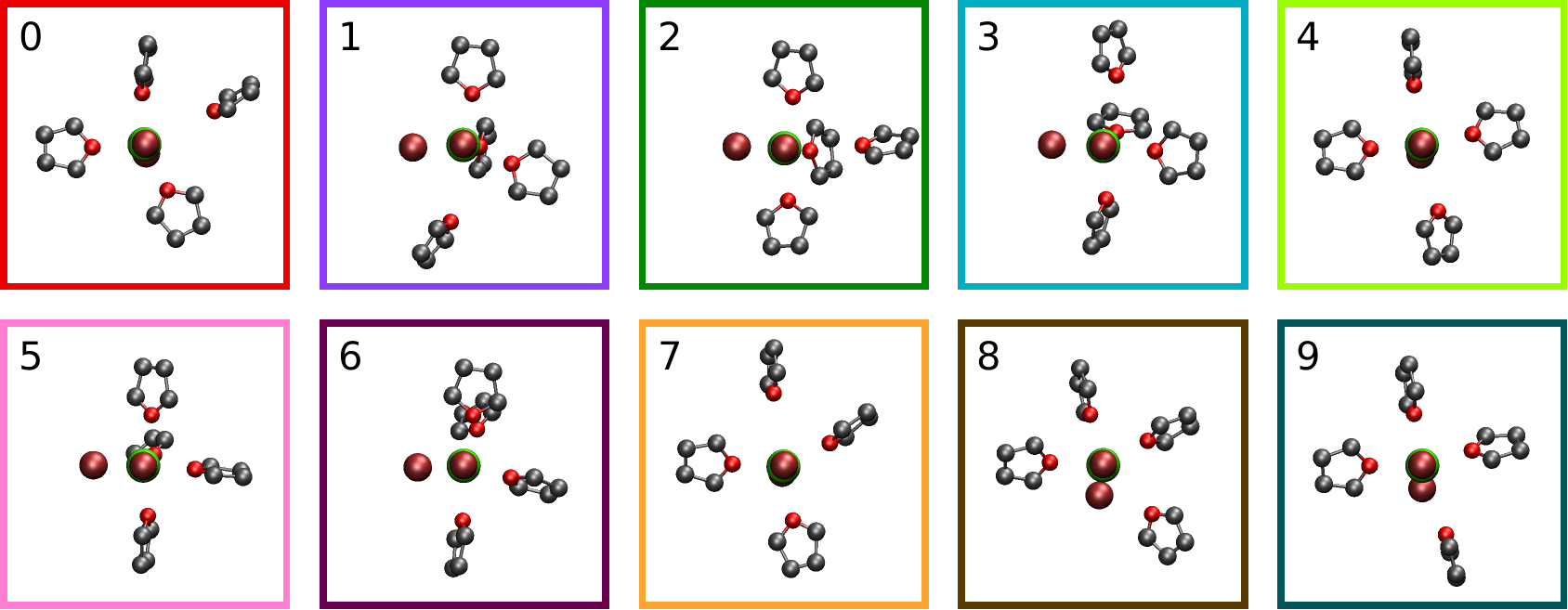}
        }}
        
    \subfloat[Side view\label{subfig:TS_C16B2CaO4_MS_side}]{
        \centerline{
        \includegraphics[width=\columnwidth]{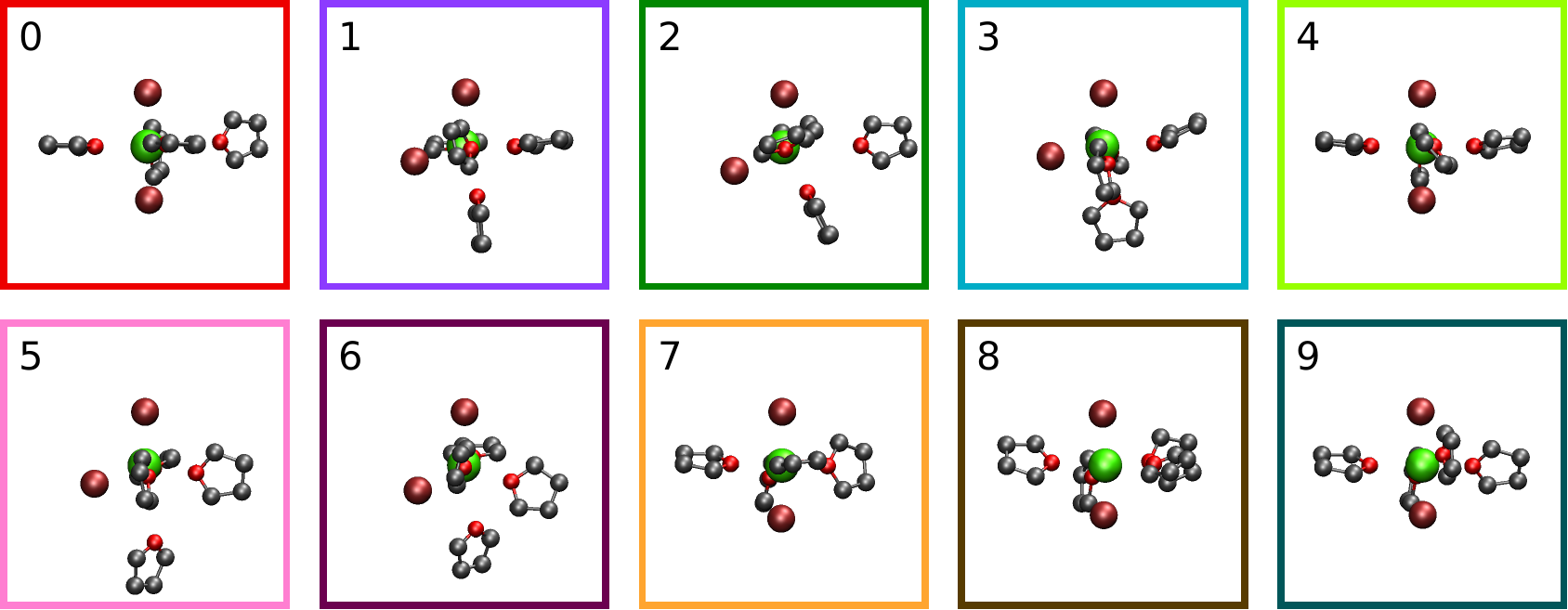}
        }}

    \caption{(a) UMAP projection colored according to HDBSCAN labelling of \ce{C16B2CaO4} local atomic arrangements of trajectory 5. Points classified as noise are labelled -1 (black). (b) Top and (c) side view of the averaged local atomic arrangements with chemical formula \ce{C16B2CaO4} for each HDBSCAN group.}
    \label{fig:TS_C16B2CaO4_UMAP_and_mean_structures}
\end{figure}

\subsection*{Nonequilibrium Sampling}

In a second example, we analyze the composition of the umbrella sampled trajectory 5 from Figure~\ref{subfig:metadyn_coordO_dZ}. This particular region in the collective variables space is at a transition point close to the graphite surface between two local free energy minima with Ca-O coordination numbers of 3 and 4, respectively. This region has a free energy only  $\sim$~5~kT higher than the lowest of the two minima (at a coordination number of 3). From the radial distribution function around calcium shown in Figure~\ref{subfig:RDF_plots} (right) we see that similarly to the previous case there are two narrow peaks at 2.35~\AA{} and 2.45~\AA{} for oxygen and boron respectively. Additionally, in the Ca-O RDF there is a second peak with lower intensity centered at 3.75~\AA{}. The second peak comes with no surprise: by constraining the coordination number to be approximately 3.5, we effectively apply a bias to the system that forces one of the THF molecules farther away from the calcium. In this case, three THF molecules maintain a distance similar to the configuration with coordination number of 4, while the fourth is pushed away, giving rise to the peak at 3.75~\AA{}. To include this second peak in our analysis, we decided to set the cutoff radius of the local atomic arrangement at 4.5~\AA{}. After isolating the environment around the calcium, we find two distinct chemical formulas: 70.4\% of the structures are \ce{C16B2CaO4} and 29.6\% are \ce{C12B2CaO3}. When multiple chemical compositions are found, a separate clustering procedure is required for each.\\

Starting from the local atomic arrangements with 3 coordinating THF molecules, we identify 7 different HDBSCAN groups (Figure~\ref{fig:TS_C12B2CaO3_UMAP_and_mean_structures}). Only groups 2 and 3, 24.1\% of the total population, have a local atomic arrangement with two axially aligned boron atoms, while the majority have a bent boron arrangement. This trend is opposite to what has been observed until now, where the majority of the local atomic arrangements had the two boron  atoms on opposite sides of the calcium ion. Again, we see how UMAP and HDBSCAN identify structures with different orientations of the THF molecules.

The structures with 4 coordinating THF molecules were divided into 10 groups and are shown in Figure~\ref{fig:TS_C16B2CaO4_UMAP_and_mean_structures}. Also in this case, there is a higher percentage of arrangements with adjacent B atoms with respect to trajectories 1-4, as 40.4 \% of the local atomic arrangements belong to groups 1, 2, 3, 5 and 6 that show the bent configuration.

Due to the fact that trajectory 5 was sampled with a biasing potential that forces the calcium-oxygen coordination number to be 3.5, the four coordinating THF molecules cannot be at the same distance from the calcium as in trajectories 1-4. Indeed, from the averaged local atomic arrangements we notice how in all cases only one THF molecule is farther away than the others. This is what gives rise to the second peak we observed in the Ca-O RDF plot in Figure~\ref{subfig:RDF_plots}. This also is a strong indication that for the system it is energetically favourable to have three THF molecules closer and only one farther away rather than multiple.\\

\begin{figure}[ht]
    \centering
    \subfloat[UMAP projections\label{subfig:m13_on_traj1234}]{
        \centerline{
        \includegraphics[width=\columnwidth]{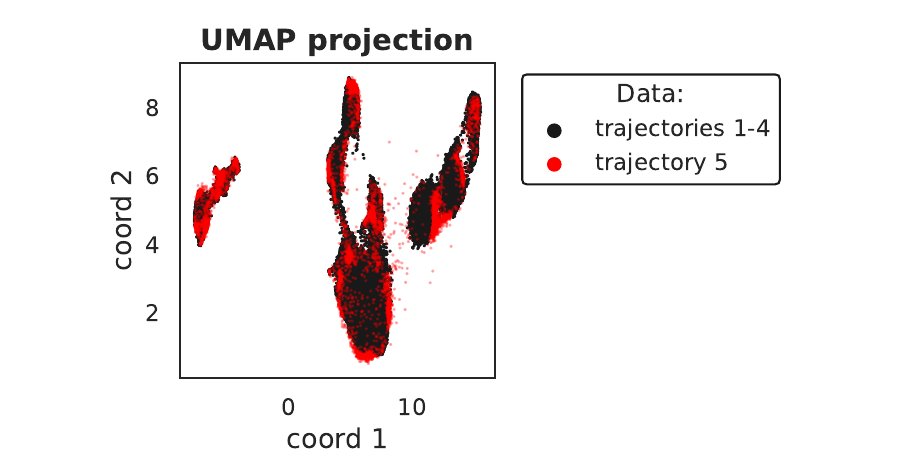}
        }}
        
    \subfloat[UMAP comparison\label{subfig:comparison_t5_to_t1-4}]{
        \centerline{
        \includegraphics[width=\columnwidth]{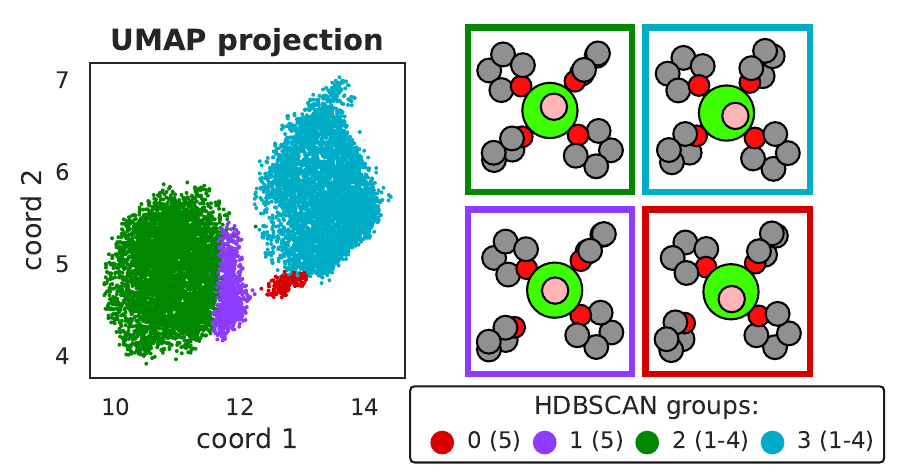}
        }}

    \caption{(a) Projection of the data from trajectory 5 on the UMAP pipeline of trajectories 1-4. Most of the newly projected data lies at the ridges or between previously existing clusters. (b)~Overlap of the UMAP projection of groups 2 and 3 from trajectories 1-4 (Figure~\ref{subfig:C16B2CaO4_UMAP}) and UMAP projection of data from trajectory 5 that is close to it and has been divided by HDBSCAN in two groups (0 and 1).}
    \label{fig:TS_new_data}
\end{figure}

In Figure~\ref{subfig:m13_on_traj1234} we show the projection of the data from the transition state on the same UMAP embedding used to cluster the data from trajectories 1-4 shown in Figure~\ref{subfig:C16B2CaO4_UMAP}. This was done by aligning the local atomic arrangements of trajectory 5 to those constituting the alignment basis set of trajectories 1-4 and then using the distance/flattened distance matrix as input to the trained UMAP pipeline. Interestingly, most of the newly projected data lies at the edges of the existing clusters or in between clusters. This indicates that the structures found in trajectory 5 are not a complete departure from points 1-4. On the other hand, the additional data could be used to better understand the link between transition states and stable states found in the MD trajectory. As an example, we focus on the data that has been projected between groups 2 and 3 in Figure~\ref{subfig:C16B2CaO4_UMAP}. Isolating those data and clustering them with HDBSCAN reveals two separate groups, one closer to group 3 and one closer to group 2. The overlap of the former projections (groups 2 and 3 from trajectory 1-4) with the new data is shown in Figure~\ref{subfig:comparison_t5_to_t1-4}.

A snapshot of the representative structures of each group is shown at the right of the UMAP plot. We clearly see from the two bottom structures that there is one THF molecule farther away from the calcium, as expected by the umbrella sampling biasing potential in trajectory 5. Moreover, we notice how structures belonging to groups that lie closer on the UMAP space have more structural similarity. Group 1 indeed looks like group 2 with an elongated THF and group 0 is closer to group 3. The RMSD value between the representative structures supports this finding:

\begin{table}[h]
\small
  \caption{\ RMSD values (in \AA{}) between the four structures in Figure~\ref{subfig:comparison_t5_to_t1-4}}
  \label{tbl:rmsd_comparison}
  \begin{tabular*}{0.48\textwidth}{@{\extracolsep{\fill}}lcccc}
    \hline
    & group 0 & group 1 & group 2 & group 3 \\
    \hline
    group 0 & 0.000 & 0.530 & 0.450 & 0.693 \\
    group 1 & 0.530 & 0.000 & 0.581 & 0.445 \\
    group 2 & 0.450 & 0.581 & 0.000 & 0.440 \\
    group 3 & 0.693 & 0.445 & 0.440 & 0.000 \\
    \hline
  \end{tabular*}
\end{table}

%% file: Sections/05_outlook_discussion.tex
\section*{Outlook and Discussion}

The main power of this unsupervised classification of local atomic arrangements is the great help it provides to the researcher to perform further analysis. Since each local atomic arrangement has been aligned to its representative and the representatives to one another, effective visualization of the atomic structures is easy. Due to the alignment, we can compute the mean structure of each group by averaging the atomic positions within each group. Although the averaged atomic positions may not have physical meaning (the bond lengths may not be maintained, for example), it is helpful for visualization purposes. And, as we  have seen, the partition of the sampled data into clusters that exhibit normal distributions for key structural characteristics indicates that the mean structure may indeed be physical. 

If the isolated clusters do emerge as normal distributions, then a promising future direction would be to explore the principal components of the variance in these distributions. Methods such as principal component analysis (PCA, for example implemented in the scikit-learn Python package~\cite{scikit-learn}) can reveal which vibrational modes are dominant in each group and perhaps serve as input for a reduced dimensional model of the system free energy with temperature dependent vibrational entropic components. 

The mean or other representatives (such as exemplars) of each group can be used as inputs for more expensive calculations of observables that might reference experiment. For example, \emph{ab-initio} electronic structure calculations can provide details relevant to both vibrational, NMR, and X-ray spectroscopy. The additional ease with which we can understand subtle variations in the distribution of representative structures as a function of the local environment (distance from an interface, for example) can also be used in complex interfacial studies to interpret differences in measurements that are specifically related to the interfacial environment -- be they structural, chemical, or merely orientational in origin.

The choice of alignment algorithm within our method is not a limitation. For the purposes of this work, we mainly used FASTOVERLAP with its Branch and Bound method for structural alignment including permutations and inversion. However, because of the the modular structure of the code, other alignment algorithms could easily be included. The shape matching IRA method was also tested \cite{Gunde2021}. In our experience, FASTOVERLAP has proven to be more reliable in correctly aligning the structures, with little to no misalignment. With IRA on the other hand we observed occasionally a misalignment to the reference structure. The probability of failure of IRA increases when the structures are not exactly or nearly congruent. Therefore, having a complete sample of reference structures that we try to align to is crucial to ensure that every coordination structure has a near-congruent representative. The best alignment results and a good trade-off in computational performance were obtained with the Branch and Bound alignment algorithm of FASTOVERLAP.

Aligning the structures to a subset of representatives instead of calculating a full distance-distance matrix has the advantage of reducing the size of the matrix that needs to be calculated. Instead of performing a $N\times N$ alignment problem (with $N$ being the total number of structures in the sample), we do a $N\times n$ alignment (with $n$ being the much smaller number of representatives). With a good choice of representatives, this reduced dataset is normally enough to differentiate the main features of the coordination environment. This leads to the fast and reliable post-processing capabilities of our algorithm. On easily available computing resources, for example a regular desktop workstation, analyzing tens of thousands of local atomic arrangements was done within minutes. The rapidity at which it is possible to obtain insights on large MD datasets further promotes an iterative approach to the analysis, where hidden relevant metrics are learned on the fly.

The advantages of this methodology are not limited merely to unsupervised dimensionality reduction. By providing previously calculated labels, for example, from the clustering of a training set, it is possible to perform supervised dimensionality reduction and metric learning. We can use this approach to train a UMAP model on a smaller or more diverse data set (maybe the metadynamics trajectory), and use the model to calculate the reduced coordinates of new structures. Once the learned and supervised metric reduction is done, the new data is given to HDBSCAN and labelled according to the previous labels. Ideally, this can be implemented into a pipeline to obtain on-the-fly classification of the coordinating environment as the simulation goes on. In the worst case, if new data presents examples that cannot be labeled -- the entirety of the collective variable space may not have been sampled in the training set -- then we are alerted to this deficiency and we can retrain with this new data to augment the number of distinct classes. We demonstrate the efficacy of the learning method on the same trajectory of the example with \ce{B2CaO4} presented above. Out of the 48,000 structures of the four umbrella sampling trajectory, we randomly took 3,000 as a training set. The clustering procedure recognized the three main coordination groups observed in Figure~\ref{subfig:B2CaO4_UMAP}. By saving a fitted dimensionality reduction pipeline, composed by normal scaling of the input coordinates and reduction of the space by UMAP, we can apply it to the "new" data composed by the 45,000 remaining clusters. The trained model was able to correctly identify 99.5\% of the new structures provided. The mislabelling only happened between the two coordination environment where the boron is on opposite sides and only the O-Ca-O angle is different. This example shows the strength of the analysis method, where less than 10\% of the data can be used to label the remaining 90\%.

%% file: Sections/06_conclusions.tex
\FloatBarrier
\section*{Conclusions}

In this work, we show how novel dimensionality reduction and hierarchical clustering algorithms can be embedded into a workflow to provide an unbiased description of coordination structure obtained through MD sampling. We isolate, classify and visualize the relevant aspects of the coordination environment of \ce{Ca^{2+}} ions solvated in an electrolyte comprising \ce{BH_4^-} anions and THF solvent in the vicinity of a graphite interface. We were able to identify the most prevalent conformational isomers of the first solvation shell of \ce{Ca^{2+}} and quantitatively estimate their population as a function of the distance from the graphite interface. The method is trivially expandable to larger solvation environments, that can reveal details on electrolyte performance \cite{Hahn2022}.

The development of new methods to process increasingly complex data sets with as little human intervention as possible is crucial in the rapidly expanding fields of high-throughput computational materials science and chemistry. Dimensionality reduction and hierarchical clustering algorithms proved to be effective tools to facilitate detailed structural analysis and to partition atomic arrangements into distinct structural distributions, which are more amenable to standard statistical analyses and comparison with experiment.

The efficient extraction of representative structural motifs with distinct coordination environments and their associated variance can greatly accelerate the interpretation of various experimental techniques with particular sensitivity to local structure and chemistry. With the approach outlined in this work, measured spectra, and their simulated analogues, may more easily be deconvoluted into contributions from specific coordination environments or regional contributions, especially from the vicinity of active interfaces. Analysis of existing molecular dynamics sampling data for the interpretation of NMR chemical shifts \cite{Markwick2010} or X-ray absorption spectroscopy -- either near-edge (NEXAFS/XANES) \cite{Prendergast2006, Wan2014} or extended fine structure (EXAFS) \cite{Fulton2010, Dang2006} -- will surely benefit, especially with recent access to operando measurements with nanometer sensitivity to chemistry at the electrode-electrolyte interface \cite{VV2014, Wu2018}. As a step in this direction, we have just applied this unsupervised learning technique to explore the formation of electroactive species in \ce{Ca(BH4)2}|THF near a graphite electrode\cite{Sanz-Matias2023}.

%% file: main.bbl
\providecommand*{\mcitethebibliography}{\thebibliography}
\csname @ifundefined\endcsname{endmcitethebibliography}
{\let\endmcitethebibliography\endthebibliography}{}
\begin{mcitethebibliography}{64}
\providecommand*{\natexlab}[1]{#1}
\providecommand*{\mciteSetBstSublistMode}[1]{}
\providecommand*{\mciteSetBstMaxWidthForm}[2]{}
\providecommand*{\mciteBstWouldAddEndPuncttrue}
  {\def\EndOfBibitem{\unskip.}}
\providecommand*{\mciteBstWouldAddEndPunctfalse}
  {\let\EndOfBibitem\relax}
\providecommand*{\mciteSetBstMidEndSepPunct}[3]{}
\providecommand*{\mciteSetBstSublistLabelBeginEnd}[3]{}
\providecommand*{\EndOfBibitem}{}
\mciteSetBstSublistMode{f}
\mciteSetBstMaxWidthForm{subitem}
{(\emph{\alph{mcitesubitemcount}})}
\mciteSetBstSublistLabelBeginEnd{\mcitemaxwidthsubitemform\space}
{\relax}{\relax}

\bibitem[Velasco-Velez \emph{et~al.}(2014)Velasco-Velez, Wu, Pascal, Wan, Guo, Prendergast, and Salmeron]{VV2014}
J.-J. Velasco-Velez, C.~H. Wu, T.~A. Pascal, L.~F. Wan, J.~Guo, D.~Prendergast and M.~Salmeron, \emph{Science}, 2014, \textbf{346}, 831--834\relax
\mciteBstWouldAddEndPuncttrue
\mciteSetBstMidEndSepPunct{\mcitedefaultmidpunct}
{\mcitedefaultendpunct}{\mcitedefaultseppunct}\relax
\EndOfBibitem
\bibitem[Sun \emph{et~al.}(2020)Sun, Yang, Ji, Zhou, Wang, Qian, and Yan]{Sun2020}
Y.~Sun, T.~Yang, H.~Ji, J.~Zhou, Z.~Wang, T.~Qian and C.~Yan, \emph{Advanced Energy Materials}, 2020, \textbf{10}, 2002373\relax
\mciteBstWouldAddEndPuncttrue
\mciteSetBstMidEndSepPunct{\mcitedefaultmidpunct}
{\mcitedefaultendpunct}{\mcitedefaultseppunct}\relax
\EndOfBibitem
\bibitem[Yang \emph{et~al.}(2020)Yang, Liu, Feng, Qian, Kao, Ha, Hahn, Seguin, Tsige, Yang, Zavadil, Persson, and Guo]{Yang2020}
F.~Yang, Y.-S. Liu, X.~Feng, K.~Qian, L.~C. Kao, Y.~Ha, N.~T. Hahn, T.~J. Seguin, M.~Tsige, W.~Yang, K.~R. Zavadil, K.~A. Persson and J.~Guo, \emph{RSC Adv.}, 2020, \textbf{10}, 27315--27321\relax
\mciteBstWouldAddEndPuncttrue
\mciteSetBstMidEndSepPunct{\mcitedefaultmidpunct}
{\mcitedefaultendpunct}{\mcitedefaultseppunct}\relax
\EndOfBibitem
\bibitem[Yamijala \emph{et~al.}(2021)Yamijala, Kwon, Guo, and Wong]{Yamijala2021}
S.~S. Yamijala, H.~Kwon, J.~Guo and B.~M. Wong, \emph{ACS Applied Materials and Interfaces}, 2021, \textbf{13}, 13114--13122\relax
\mciteBstWouldAddEndPuncttrue
\mciteSetBstMidEndSepPunct{\mcitedefaultmidpunct}
{\mcitedefaultendpunct}{\mcitedefaultseppunct}\relax
\EndOfBibitem
\bibitem[Young and Smeu(2021)]{Young2021}
J.~Young and M.~Smeu, \emph{Advanced Theory and Simulations}, 2021, \textbf{4}, 2100018\relax
\mciteBstWouldAddEndPuncttrue
\mciteSetBstMidEndSepPunct{\mcitedefaultmidpunct}
{\mcitedefaultendpunct}{\mcitedefaultseppunct}\relax
\EndOfBibitem
\bibitem[Yao \emph{et~al.}(2022)Yao, Chen, Fu, and Zhang]{Yao2022}
N.~Yao, X.~Chen, Z.~H. Fu and Q.~Zhang, \emph{Chemical Reviews}, 2022, \textbf{122}, 10970--11021\relax
\mciteBstWouldAddEndPuncttrue
\mciteSetBstMidEndSepPunct{\mcitedefaultmidpunct}
{\mcitedefaultendpunct}{\mcitedefaultseppunct}\relax
\EndOfBibitem
\bibitem[McInnes \emph{et~al.}(2018)McInnes, Healy, and Melville]{McInnes2018}
L.~McInnes, J.~Healy and J.~Melville, \emph{UMAP: Uniform Manifold Approximation and Projection for Dimension Reduction}, 2018, \url{https://arxiv.org/abs/1802.03426}\relax
\mciteBstWouldAddEndPuncttrue
\mciteSetBstMidEndSepPunct{\mcitedefaultmidpunct}
{\mcitedefaultendpunct}{\mcitedefaultseppunct}\relax
\EndOfBibitem
\bibitem[McInnes \emph{et~al.}(2017)McInnes, Healy, and Astels]{McInnes2017}
L.~McInnes, J.~Healy and S.~Astels, \emph{Journal of Open Source Software}, 2017, \textbf{2}, 205\relax
\mciteBstWouldAddEndPuncttrue
\mciteSetBstMidEndSepPunct{\mcitedefaultmidpunct}
{\mcitedefaultendpunct}{\mcitedefaultseppunct}\relax
\EndOfBibitem
\bibitem[Griffiths \emph{et~al.}(2017)Griffiths, Niblett, and Wales]{Griffiths2017}
M.~Griffiths, S.~P. Niblett and D.~J. Wales, \emph{Journal of Chemical Theory and Computation}, 2017, \textbf{13}, 4914--4931\relax
\mciteBstWouldAddEndPuncttrue
\mciteSetBstMidEndSepPunct{\mcitedefaultmidpunct}
{\mcitedefaultendpunct}{\mcitedefaultseppunct}\relax
\EndOfBibitem
\bibitem[Thompson \emph{et~al.}(2022)Thompson, Aktulga, Berger, Bolintineanu, Brown, Crozier, in~'t Veld, Kohlmeyer, Moore, Nguyen, Shan, Stevens, Tranchida, Trott, and Plimpton]{LAMMPS}
A.~P. Thompson, H.~M. Aktulga, R.~Berger, D.~S. Bolintineanu, W.~M. Brown, P.~S. Crozier, P.~J. in~'t Veld, A.~Kohlmeyer, S.~G. Moore, T.~D. Nguyen, R.~Shan, M.~J. Stevens, J.~Tranchida, C.~Trott and S.~J. Plimpton, \emph{Comp. Phys. Comm.}, 2022, \textbf{271}, 108171\relax
\mciteBstWouldAddEndPuncttrue
\mciteSetBstMidEndSepPunct{\mcitedefaultmidpunct}
{\mcitedefaultendpunct}{\mcitedefaultseppunct}\relax
\EndOfBibitem
\bibitem[Torrie and Valleau(1977)]{Torrie1977}
G.~M. Torrie and J.~P. Valleau, \emph{Journal of Computational Physics}, 1977, \textbf{23}, 187--199\relax
\mciteBstWouldAddEndPuncttrue
\mciteSetBstMidEndSepPunct{\mcitedefaultmidpunct}
{\mcitedefaultendpunct}{\mcitedefaultseppunct}\relax
\EndOfBibitem
\bibitem[Laio and Parrinello(2002)]{Laio2002}
A.~Laio and M.~Parrinello, \emph{Proceedings of the National Academy of Sciences of the United States of America}, 2002, \textbf{99}, 12562\relax
\mciteBstWouldAddEndPuncttrue
\mciteSetBstMidEndSepPunct{\mcitedefaultmidpunct}
{\mcitedefaultendpunct}{\mcitedefaultseppunct}\relax
\EndOfBibitem
\bibitem[Fiorin \emph{et~al.}(2013)Fiorin, Klein, and Hénin]{Fiorin2013}
G.~Fiorin, M.~L. Klein and J.~Hénin, \emph{Molecular Physics}, 2013, \textbf{111}, 3345--3362\relax
\mciteBstWouldAddEndPuncttrue
\mciteSetBstMidEndSepPunct{\mcitedefaultmidpunct}
{\mcitedefaultendpunct}{\mcitedefaultseppunct}\relax
\EndOfBibitem
\bibitem[Bussi and Laio(2020)]{Bussi2020}
G.~Bussi and A.~Laio, \emph{Nature Reviews Physics}, 2020, \textbf{2}, 200--212\relax
\mciteBstWouldAddEndPuncttrue
\mciteSetBstMidEndSepPunct{\mcitedefaultmidpunct}
{\mcitedefaultendpunct}{\mcitedefaultseppunct}\relax
\EndOfBibitem
\bibitem[Baskin and Prendergast(2019)]{Baskin2019}
A.~Baskin and D.~Prendergast, \emph{Journal of Physical Chemistry Letters}, 2019, \textbf{10}, 4920--4928\relax
\mciteBstWouldAddEndPuncttrue
\mciteSetBstMidEndSepPunct{\mcitedefaultmidpunct}
{\mcitedefaultendpunct}{\mcitedefaultseppunct}\relax
\EndOfBibitem
\bibitem[Roy \emph{et~al.}(2016)Roy, Baer, Mundy, and Schenter]{Roy2016}
S.~Roy, M.~D. Baer, C.~J. Mundy and G.~K. Schenter, \emph{Journal of Physical Chemistry C}, 2016, \textbf{120}, 7597--7605\relax
\mciteBstWouldAddEndPuncttrue
\mciteSetBstMidEndSepPunct{\mcitedefaultmidpunct}
{\mcitedefaultendpunct}{\mcitedefaultseppunct}\relax
\EndOfBibitem
\bibitem[Byrne \emph{et~al.}(2017)Byrne, Raiteri, and Gale]{Byrne2017}
E.~H. Byrne, P.~Raiteri and J.~D. Gale, \emph{Journal of Physical Chemistry C}, 2017, \textbf{121}, 25956--25966\relax
\mciteBstWouldAddEndPuncttrue
\mciteSetBstMidEndSepPunct{\mcitedefaultmidpunct}
{\mcitedefaultendpunct}{\mcitedefaultseppunct}\relax
\EndOfBibitem
\bibitem[Camacho-Forero \emph{et~al.}(2015)Camacho-Forero, Smith, Bertolini, and Balbuena]{Camacho2015}
L.~E. Camacho-Forero, T.~W. Smith, S.~Bertolini and P.~B. Balbuena, \emph{Journal of Physical Chemistry C}, 2015, \textbf{119}, 26828--26839\relax
\mciteBstWouldAddEndPuncttrue
\mciteSetBstMidEndSepPunct{\mcitedefaultmidpunct}
{\mcitedefaultendpunct}{\mcitedefaultseppunct}\relax
\EndOfBibitem
\bibitem[Nandasiri \emph{et~al.}(2017)Nandasiri, Camacho-Forero, Schwarz, Shutthanandan, Thevuthasan, Balbuena, Mueller, and Murugesan]{Nandasiri2017}
M.~I. Nandasiri, L.~E. Camacho-Forero, A.~M. Schwarz, V.~Shutthanandan, S.~Thevuthasan, P.~B. Balbuena, K.~T. Mueller and V.~Murugesan, \emph{Chemistry of Materials}, 2017, \textbf{29}, 4728--4737\relax
\mciteBstWouldAddEndPuncttrue
\mciteSetBstMidEndSepPunct{\mcitedefaultmidpunct}
{\mcitedefaultendpunct}{\mcitedefaultseppunct}\relax
\EndOfBibitem
\bibitem[Hu \emph{et~al.}(2018)Hu, Rajput, Wan, Shao, Deng, Jaegers, Hu, Chen, Shin, Monk, Chen, Qin, Mueller, Liu, and Persson]{Hu2018}
J.~Z. Hu, N.~N. Rajput, C.~Wan, Y.~Shao, X.~Deng, N.~R. Jaegers, M.~Hu, Y.~Chen, Y.~Shin, J.~Monk, Z.~Chen, Z.~Qin, K.~T. Mueller, J.~Liu and K.~A. Persson, \emph{Nano Energy}, 2018, \textbf{46}, 436--446\relax
\mciteBstWouldAddEndPuncttrue
\mciteSetBstMidEndSepPunct{\mcitedefaultmidpunct}
{\mcitedefaultendpunct}{\mcitedefaultseppunct}\relax
\EndOfBibitem
\bibitem[Hahn \emph{et~al.}(2020)Hahn, Self, Seguin, Driscoll, Rodriguez, Balasubramanian, Persson, and Zavadil]{Hahn2020}
N.~T. Hahn, J.~Self, T.~J. Seguin, D.~M. Driscoll, M.~A. Rodriguez, M.~Balasubramanian, K.~A. Persson and K.~R. Zavadil, \emph{Journal of Materials Chemistry A}, 2020, \textbf{8}, 7235--7244\relax
\mciteBstWouldAddEndPuncttrue
\mciteSetBstMidEndSepPunct{\mcitedefaultmidpunct}
{\mcitedefaultendpunct}{\mcitedefaultseppunct}\relax
\EndOfBibitem
\bibitem[Kao \emph{et~al.}(2020)Kao, Feng, Ha, Yang, Liu, Hahn, MacDougall, Chao, Yang, Zavadil, and Guo]{Kao2020}
L.~C. Kao, X.~Feng, Y.~Ha, F.~Yang, Y.-S. Liu, N.~T. Hahn, J.~MacDougall, W.~Chao, W.~Yang, K.~R. Zavadil and J.~Guo, \emph{Surface Science}, 2020, \textbf{702}, 121720\relax
\mciteBstWouldAddEndPuncttrue
\mciteSetBstMidEndSepPunct{\mcitedefaultmidpunct}
{\mcitedefaultendpunct}{\mcitedefaultseppunct}\relax
\EndOfBibitem
\bibitem[Agarwal \emph{et~al.}(2021)Agarwal, Howard, Prabhakaran, Johnson, Murugesan, Mueller, Curtiss, and Assary]{Agarwal2021}
G.~Agarwal, J.~D. Howard, V.~Prabhakaran, G.~E. Johnson, V.~Murugesan, K.~T. Mueller, L.~A. Curtiss and R.~S. Assary, \emph{ACS Applied Materials and Interfaces}, 2021, \textbf{13}, 38816--38825\relax
\mciteBstWouldAddEndPuncttrue
\mciteSetBstMidEndSepPunct{\mcitedefaultmidpunct}
{\mcitedefaultendpunct}{\mcitedefaultseppunct}\relax
\EndOfBibitem
\bibitem[Xu(2004)]{Xu2004}
K.~Xu, \emph{Chemical Reviews}, 2004, \textbf{104}, 4303--4417\relax
\mciteBstWouldAddEndPuncttrue
\mciteSetBstMidEndSepPunct{\mcitedefaultmidpunct}
{\mcitedefaultendpunct}{\mcitedefaultseppunct}\relax
\EndOfBibitem
\bibitem[Kerisit \emph{et~al.}(2006)Kerisit, Ilton, and Parker]{Kerisit2006}
S.~Kerisit, E.~S. Ilton and S.~C. Parker, \emph{Journal of Physical Chemistry B}, 2006, \textbf{110}, 20491--20501\relax
\mciteBstWouldAddEndPuncttrue
\mciteSetBstMidEndSepPunct{\mcitedefaultmidpunct}
{\mcitedefaultendpunct}{\mcitedefaultseppunct}\relax
\EndOfBibitem
\bibitem[Bedrov \emph{et~al.}(2019)Bedrov, Piquemal, Borodin, MacKerell, Roux, and Schröder]{Bedrov2019}
D.~Bedrov, J.~P. Piquemal, O.~Borodin, A.~D. MacKerell, B.~Roux and C.~Schröder, \emph{Chemical Reviews}, 2019, \textbf{119}, 7940--7995\relax
\mciteBstWouldAddEndPuncttrue
\mciteSetBstMidEndSepPunct{\mcitedefaultmidpunct}
{\mcitedefaultendpunct}{\mcitedefaultseppunct}\relax
\EndOfBibitem
\bibitem[Pham \emph{et~al.}(2017)Pham, Kweon, Samanta, Lordi, and Pask]{Pham2017}
T.~A. Pham, K.~E. Kweon, A.~Samanta, V.~Lordi and J.~E. Pask, \emph{Journal of Physical Chemistry C}, 2017, \textbf{121}, 21913--21920\relax
\mciteBstWouldAddEndPuncttrue
\mciteSetBstMidEndSepPunct{\mcitedefaultmidpunct}
{\mcitedefaultendpunct}{\mcitedefaultseppunct}\relax
\EndOfBibitem
\bibitem[Rajput \emph{et~al.}(2018)Rajput, Seguin, Wood, Qu, and Persson]{Rajput2018}
N.~N. Rajput, T.~J. Seguin, B.~M. Wood, X.~Qu and K.~A. Persson, \emph{Topics in Current Chemistry}, 2018, \textbf{376}, 19\relax
\mciteBstWouldAddEndPuncttrue
\mciteSetBstMidEndSepPunct{\mcitedefaultmidpunct}
{\mcitedefaultendpunct}{\mcitedefaultseppunct}\relax
\EndOfBibitem
\bibitem[Han(2019)]{Han2019}
S.~Han, \emph{Scientific Reports}, 2019, \textbf{9}, 5555\relax
\mciteBstWouldAddEndPuncttrue
\mciteSetBstMidEndSepPunct{\mcitedefaultmidpunct}
{\mcitedefaultendpunct}{\mcitedefaultseppunct}\relax
\EndOfBibitem
\bibitem[Bezabh \emph{et~al.}(2020)Bezabh, Tsai, Hagos, Beyene, Berhe, Hagos, Abrha, Chiu, Su, and Hwang]{Bezabh2020}
H.~K. Bezabh, M.~C. Tsai, T.~T. Hagos, T.~T. Beyene, G.~B. Berhe, T.~M. Hagos, L.~H. Abrha, S.~F. Chiu, W.~N. Su and B.~J. Hwang, \emph{Electrochemistry Communications}, 2020, \textbf{113}, 106685\relax
\mciteBstWouldAddEndPuncttrue
\mciteSetBstMidEndSepPunct{\mcitedefaultmidpunct}
{\mcitedefaultendpunct}{\mcitedefaultseppunct}\relax
\EndOfBibitem
\bibitem[Hou \emph{et~al.}(2021)Hou, Fong, Wang, and Persson]{Hou2021}
T.~Hou, K.~D. Fong, J.~Wang and K.~A. Persson, \emph{Chemical Science}, 2021, \textbf{12}, 14740--14751\relax
\mciteBstWouldAddEndPuncttrue
\mciteSetBstMidEndSepPunct{\mcitedefaultmidpunct}
{\mcitedefaultendpunct}{\mcitedefaultseppunct}\relax
\EndOfBibitem
\bibitem[Tian \emph{et~al.}(2022)Tian, Zou, Liu, Wang, Yin, Ming, and Alshareef]{Tian2022}
Z.~Tian, Y.~Zou, G.~Liu, Y.~Wang, J.~Yin, J.~Ming and H.~N. Alshareef, \emph{Advanced Science}, 2022, \textbf{9}, 2201207\relax
\mciteBstWouldAddEndPuncttrue
\mciteSetBstMidEndSepPunct{\mcitedefaultmidpunct}
{\mcitedefaultendpunct}{\mcitedefaultseppunct}\relax
\EndOfBibitem
\bibitem[Konstantinovsky \emph{et~al.}(2022)Konstantinovsky, Perets, Santiago, Velarde, Hammes-Schiffer, and Yan]{Konstantinovsky2022}
D.~Konstantinovsky, E.~A. Perets, T.~Santiago, L.~Velarde, S.~Hammes-Schiffer and E.~C. Yan, \emph{ACS Central Science}, 2022, \textbf{8}, 1404--1414\relax
\mciteBstWouldAddEndPuncttrue
\mciteSetBstMidEndSepPunct{\mcitedefaultmidpunct}
{\mcitedefaultendpunct}{\mcitedefaultseppunct}\relax
\EndOfBibitem
\bibitem[Yu \emph{et~al.}(2022)Yu, Juran, Liu, Han, Wang, Mueller, Ma, Xu, Li, Curtiss, and Cheng]{Yu2022}
Z.~Yu, T.~R. Juran, X.~Liu, K.~S. Han, H.~Wang, K.~T. Mueller, L.~Ma, K.~Xu, T.~Li, L.~A. Curtiss and L.~Cheng, \emph{Energy and Environmental Materials}, 2022, \textbf{5}, 295--304\relax
\mciteBstWouldAddEndPuncttrue
\mciteSetBstMidEndSepPunct{\mcitedefaultmidpunct}
{\mcitedefaultendpunct}{\mcitedefaultseppunct}\relax
\EndOfBibitem
\bibitem[Kufareva and Abagyan(2012)]{Kufareva2012}
I.~Kufareva and R.~Abagyan, \emph{Methods in Molecular Biology}, 2012, \textbf{857}, 231--257\relax
\mciteBstWouldAddEndPuncttrue
\mciteSetBstMidEndSepPunct{\mcitedefaultmidpunct}
{\mcitedefaultendpunct}{\mcitedefaultseppunct}\relax
\EndOfBibitem
\bibitem[Rupp \emph{et~al.}(2012)Rupp, Tkatchenko, M\"uller, and von Lilienfeld]{Rupp2012}
M.~Rupp, A.~Tkatchenko, K.-R. M\"uller and O.~A. von Lilienfeld, \emph{Phys. Rev. Lett.}, 2012, \textbf{108}, 058301\relax
\mciteBstWouldAddEndPuncttrue
\mciteSetBstMidEndSepPunct{\mcitedefaultmidpunct}
{\mcitedefaultendpunct}{\mcitedefaultseppunct}\relax
\EndOfBibitem
\bibitem[Hansen \emph{et~al.}(2015)Hansen, Biegler, Ramakrishnan, Pronobis, Lilienfeld, Müller, and Tkatchenko]{Hansen2015}
K.~Hansen, F.~Biegler, R.~Ramakrishnan, W.~Pronobis, O.~A.~V. Lilienfeld, K.~R. Müller and A.~Tkatchenko, \emph{Journal of Physical Chemistry Letters}, 2015, \textbf{6}, 2326--2331\relax
\mciteBstWouldAddEndPuncttrue
\mciteSetBstMidEndSepPunct{\mcitedefaultmidpunct}
{\mcitedefaultendpunct}{\mcitedefaultseppunct}\relax
\EndOfBibitem
\bibitem[Elton \emph{et~al.}(2018)Elton, Boukouvalas, Butrico, Fuge, and Chung]{Elton2018}
D.~C. Elton, Z.~Boukouvalas, M.~S. Butrico, M.~D. Fuge and P.~W. Chung, \emph{Scientific Reports}, 2018, \textbf{8}, 9059\relax
\mciteBstWouldAddEndPuncttrue
\mciteSetBstMidEndSepPunct{\mcitedefaultmidpunct}
{\mcitedefaultendpunct}{\mcitedefaultseppunct}\relax
\EndOfBibitem
\bibitem[Ceriotti \emph{et~al.}(2018)Ceriotti, Willatt, and Cs{\'a}nyi]{Ceriotti2018}
M.~Ceriotti, M.~J. Willatt and G.~Cs{\'a}nyi, in \emph{Machine Learning of Atomic-Scale Properties Based on Physical Principles}, ed. W.~Andreoni and S.~Yip, Springer International Publishing, Cham, 2018, pp. 1--27\relax
\mciteBstWouldAddEndPuncttrue
\mciteSetBstMidEndSepPunct{\mcitedefaultmidpunct}
{\mcitedefaultendpunct}{\mcitedefaultseppunct}\relax
\EndOfBibitem
\bibitem[Reinhart(2021)]{Reinhart2021}
W.~F. Reinhart, \emph{Computational Materials Science}, 2021, \textbf{196}, 110511\relax
\mciteBstWouldAddEndPuncttrue
\mciteSetBstMidEndSepPunct{\mcitedefaultmidpunct}
{\mcitedefaultendpunct}{\mcitedefaultseppunct}\relax
\EndOfBibitem
\bibitem[Kuhn(1955)]{Kuhn1955}
H.~W. Kuhn, \emph{Naval Research Logistics Quarterly}, 1955, \textbf{2}, 83--97\relax
\mciteBstWouldAddEndPuncttrue
\mciteSetBstMidEndSepPunct{\mcitedefaultmidpunct}
{\mcitedefaultendpunct}{\mcitedefaultseppunct}\relax
\EndOfBibitem
\bibitem[Jonker and Volgenant(1987)]{Jonker1987}
R.~Jonker and A.~Volgenant, \emph{Computing}, 1987, \textbf{38}, 325--340\relax
\mciteBstWouldAddEndPuncttrue
\mciteSetBstMidEndSepPunct{\mcitedefaultmidpunct}
{\mcitedefaultendpunct}{\mcitedefaultseppunct}\relax
\EndOfBibitem
\bibitem[Carpaneto \emph{et~al.}(1988)Carpaneto, Martello, and Toth]{Carpaneto1988}
G.~Carpaneto, S.~Martello and P.~Toth, \emph{Annals of Operations Research}, 1988, \textbf{13}, 191--223\relax
\mciteBstWouldAddEndPuncttrue
\mciteSetBstMidEndSepPunct{\mcitedefaultmidpunct}
{\mcitedefaultendpunct}{\mcitedefaultseppunct}\relax
\EndOfBibitem
\bibitem[Mart{\'\i}nez \emph{et~al.}(2009)Mart{\'\i}nez, Andrade, Birgin, and Mart{\'\i}nez]{Martinez2009}
L.~Mart{\'\i}nez, R.~Andrade, E.~G. Birgin and J.~M. Mart{\'\i}nez, \emph{J Comput Chem}, 2009, \textbf{30}, 2157--2164\relax
\mciteBstWouldAddEndPuncttrue
\mciteSetBstMidEndSepPunct{\mcitedefaultmidpunct}
{\mcitedefaultendpunct}{\mcitedefaultseppunct}\relax
\EndOfBibitem
\bibitem[Sambasivarao and Acevedo(2009)]{Samba2009}
S.~V. Sambasivarao and O.~Acevedo, \emph{Journal of Chemical Theory and Computation}, 2009, \textbf{5}, 1038--1050\relax
\mciteBstWouldAddEndPuncttrue
\mciteSetBstMidEndSepPunct{\mcitedefaultmidpunct}
{\mcitedefaultendpunct}{\mcitedefaultseppunct}\relax
\EndOfBibitem
\bibitem[Baskin and Prendergast(2020)]{Baskin2020}
A.~Baskin and D.~Prendergast, \emph{The Journal of Physical Chemistry Letters}, 2020, \textbf{11}, 9336--9343\relax
\mciteBstWouldAddEndPuncttrue
\mciteSetBstMidEndSepPunct{\mcitedefaultmidpunct}
{\mcitedefaultendpunct}{\mcitedefaultseppunct}\relax
\EndOfBibitem
\bibitem[Driscoll \emph{et~al.}(2020)Driscoll, Dandu, Hahn, Seguin, Persson, Zavadil, Curtiss, and Balasubramanian]{Driscoll2020}
D.~M. Driscoll, N.~K. Dandu, N.~T. Hahn, T.~J. Seguin, K.~A. Persson, K.~R. Zavadil, L.~A. Curtiss and M.~Balasubramanian, \emph{Journal of The Electrochemical Society}, 2020, \textbf{167}, 160512\relax
\mciteBstWouldAddEndPuncttrue
\mciteSetBstMidEndSepPunct{\mcitedefaultmidpunct}
{\mcitedefaultendpunct}{\mcitedefaultseppunct}\relax
\EndOfBibitem
\bibitem[Nosé(1984)]{Nose1984}
S.~Nosé, \emph{The Journal of Chemical Physics}, 1984, \textbf{81}, 511--519\relax
\mciteBstWouldAddEndPuncttrue
\mciteSetBstMidEndSepPunct{\mcitedefaultmidpunct}
{\mcitedefaultendpunct}{\mcitedefaultseppunct}\relax
\EndOfBibitem
\bibitem[Hoover(1985)]{Hoover1985}
W.~G. Hoover, \emph{Phys. Rev. A}, 1985, \textbf{31}, 1695--1697\relax
\mciteBstWouldAddEndPuncttrue
\mciteSetBstMidEndSepPunct{\mcitedefaultmidpunct}
{\mcitedefaultendpunct}{\mcitedefaultseppunct}\relax
\EndOfBibitem
\bibitem[Raiteri \emph{et~al.}(2006)Raiteri, Laio, Gervasio, Micheletti, and Parrinello]{Raiteri2006}
P.~Raiteri, A.~Laio, F.~L. Gervasio, C.~Micheletti and M.~Parrinello, \emph{Journal of Physical Chemistry B}, 2006, \textbf{110}, 3533--3539\relax
\mciteBstWouldAddEndPuncttrue
\mciteSetBstMidEndSepPunct{\mcitedefaultmidpunct}
{\mcitedefaultendpunct}{\mcitedefaultseppunct}\relax
\EndOfBibitem
\bibitem[Pedregosa \emph{et~al.}(2011)Pedregosa, Varoquaux, Gramfort, Michel, Thirion, Grisel, Blondel, Prettenhofer, Weiss, Dubourg, Vanderplas, Passos, Cournapeau, Brucher, Perrot, and Duchesnay]{scikit-learn}
F.~Pedregosa, G.~Varoquaux, A.~Gramfort, V.~Michel, B.~Thirion, O.~Grisel, M.~Blondel, P.~Prettenhofer, R.~Weiss, V.~Dubourg, J.~Vanderplas, A.~Passos, D.~Cournapeau, M.~Brucher, M.~Perrot and E.~Duchesnay, \emph{Journal of Machine Learning Research}, 2011, \textbf{12}, 2825--2830\relax
\mciteBstWouldAddEndPuncttrue
\mciteSetBstMidEndSepPunct{\mcitedefaultmidpunct}
{\mcitedefaultendpunct}{\mcitedefaultseppunct}\relax
\EndOfBibitem
\bibitem[Harris \emph{et~al.}(2020)Harris, Millman, van~der Walt, Gommers, Virtanen, Cournapeau, Wieser, Taylor, Berg, Smith, Kern, Picus, Hoyer, van Kerkwijk, Brett, Haldane, del R{\'{i}}o, Wiebe, Peterson, G{\'{e}}rard-Marchant, Sheppard, Reddy, Weckesser, Abbasi, Gohlke, and Oliphant]{numpy}
C.~R. Harris, K.~J. Millman, S.~J. van~der Walt, R.~Gommers, P.~Virtanen, D.~Cournapeau, E.~Wieser, J.~Taylor, S.~Berg, N.~J. Smith, R.~Kern, M.~Picus, S.~Hoyer, M.~H. van Kerkwijk, M.~Brett, A.~Haldane, J.~F. del R{\'{i}}o, M.~Wiebe, P.~Peterson, P.~G{\'{e}}rard-Marchant, K.~Sheppard, T.~Reddy, W.~Weckesser, H.~Abbasi, C.~Gohlke and T.~E. Oliphant, \emph{Nature}, 2020, \textbf{585}, 357--362\relax
\mciteBstWouldAddEndPuncttrue
\mciteSetBstMidEndSepPunct{\mcitedefaultmidpunct}
{\mcitedefaultendpunct}{\mcitedefaultseppunct}\relax
\EndOfBibitem
\bibitem[Larsen \emph{et~al.}(2017)Larsen, Mortensen, Blomqvist, Castelli, Christensen, Dułak, Friis, Groves, Hammer, Hargus, Hermes, Jennings, Jensen, Kermode, Kitchin, Kolsbjerg, Kubal, Kaasbjerg, Lysgaard, Maronsson, Maxson, Olsen, Pastewka, Peterson, Rostgaard, Schiøtz, Schütt, Strange, Thygesen, Vegge, Vilhelmsen, Walter, Zeng, and Jacobsen]{Larsen2017}
A.~H. Larsen, J.~J. Mortensen, J.~Blomqvist, I.~E. Castelli, R.~Christensen, M.~Dułak, J.~Friis, M.~N. Groves, B.~Hammer, C.~Hargus, E.~D. Hermes, P.~C. Jennings, P.~B. Jensen, J.~Kermode, J.~R. Kitchin, E.~L. Kolsbjerg, J.~Kubal, K.~Kaasbjerg, S.~Lysgaard, J.~B. Maronsson, T.~Maxson, T.~Olsen, L.~Pastewka, A.~Peterson, C.~Rostgaard, J.~Schiøtz, O.~Schütt, M.~Strange, K.~S. Thygesen, T.~Vegge, L.~Vilhelmsen, M.~Walter, Z.~Zeng and K.~W. Jacobsen, \emph{Journal of Physics: Condensed Matter}, 2017, \textbf{29}, 273002\relax
\mciteBstWouldAddEndPuncttrue
\mciteSetBstMidEndSepPunct{\mcitedefaultmidpunct}
{\mcitedefaultendpunct}{\mcitedefaultseppunct}\relax
\EndOfBibitem
\bibitem[Hunter(2007)]{matplotlib}
J.~D. Hunter, \emph{Computing in Science \& Engineering}, 2007, \textbf{9}, 90--95\relax
\mciteBstWouldAddEndPuncttrue
\mciteSetBstMidEndSepPunct{\mcitedefaultmidpunct}
{\mcitedefaultendpunct}{\mcitedefaultseppunct}\relax
\EndOfBibitem
\bibitem[Waskom(2021)]{seaborn}
M.~L. Waskom, \emph{Journal of Open Source Software}, 2021, \textbf{6}, 3021\relax
\mciteBstWouldAddEndPuncttrue
\mciteSetBstMidEndSepPunct{\mcitedefaultmidpunct}
{\mcitedefaultendpunct}{\mcitedefaultseppunct}\relax
\EndOfBibitem
\bibitem[Gunde \emph{et~al.}(2021)Gunde, Salles, Hémeryck, and Martin-Samos]{Gunde2021}
M.~Gunde, N.~Salles, A.~Hémeryck and L.~Martin-Samos, \emph{Journal of Chemical Information and Modeling}, 2021, \textbf{61}, 5446--5457\relax
\mciteBstWouldAddEndPuncttrue
\mciteSetBstMidEndSepPunct{\mcitedefaultmidpunct}
{\mcitedefaultendpunct}{\mcitedefaultseppunct}\relax
\EndOfBibitem
\bibitem[Hahn \emph{et~al.}(2022)Hahn, Self, Driscoll, Dandu, Han, Murugesan, Mueller, Curtiss, Balasubramanian, Persson, and Zavadil]{Hahn2022}
N.~T. Hahn, J.~Self, D.~M. Driscoll, N.~Dandu, K.~S. Han, V.~Murugesan, K.~T. Mueller, L.~A. Curtiss, M.~Balasubramanian, K.~A. Persson and K.~R. Zavadil, \emph{Physical Chemistry Chemical Physics}, 2022, \textbf{24}, 674--686\relax
\mciteBstWouldAddEndPuncttrue
\mciteSetBstMidEndSepPunct{\mcitedefaultmidpunct}
{\mcitedefaultendpunct}{\mcitedefaultseppunct}\relax
\EndOfBibitem
\bibitem[Markwick \emph{et~al.}(2010)Markwick, Cervantes, Abel, Komives, Blackledge, and McCammon]{Markwick2010}
P.~R. Markwick, C.~F. Cervantes, B.~L. Abel, E.~A. Komives, M.~Blackledge and J.~A. McCammon, \emph{Journal of the American Chemical Society}, 2010, \textbf{132}, 1220--1221\relax
\mciteBstWouldAddEndPuncttrue
\mciteSetBstMidEndSepPunct{\mcitedefaultmidpunct}
{\mcitedefaultendpunct}{\mcitedefaultseppunct}\relax
\EndOfBibitem
\bibitem[Prendergast and Galli(2006)]{Prendergast2006}
D.~Prendergast and G.~Galli, \emph{Physical Review Letters}, 2006, \textbf{96}, 215502\relax
\mciteBstWouldAddEndPuncttrue
\mciteSetBstMidEndSepPunct{\mcitedefaultmidpunct}
{\mcitedefaultendpunct}{\mcitedefaultseppunct}\relax
\EndOfBibitem
\bibitem[Wan and Prendergast(2014)]{Wan2014}
L.~F. Wan and D.~Prendergast, \emph{Journal of the American Chemical Society}, 2014, \textbf{136}, 14456--14464\relax
\mciteBstWouldAddEndPuncttrue
\mciteSetBstMidEndSepPunct{\mcitedefaultmidpunct}
{\mcitedefaultendpunct}{\mcitedefaultseppunct}\relax
\EndOfBibitem
\bibitem[Fulton \emph{et~al.}(2010)Fulton, Schenter, Baer, Mundy, Dang, and Balasubramanian]{Fulton2010}
J.~L. Fulton, G.~K. Schenter, M.~D. Baer, C.~J. Mundy, L.~X. Dang and M.~Balasubramanian, \emph{Journal of Physical Chemistry B}, 2010, \textbf{114}, 12926--12937\relax
\mciteBstWouldAddEndPuncttrue
\mciteSetBstMidEndSepPunct{\mcitedefaultmidpunct}
{\mcitedefaultendpunct}{\mcitedefaultseppunct}\relax
\EndOfBibitem
\bibitem[Dang \emph{et~al.}(2006)Dang, Schenter, Glezakou, and Fulton]{Dang2006}
L.~X. Dang, G.~K. Schenter, V.~A. Glezakou and J.~L. Fulton, \emph{Journal of Physical Chemistry B}, 2006, \textbf{110}, 23644--23654\relax
\mciteBstWouldAddEndPuncttrue
\mciteSetBstMidEndSepPunct{\mcitedefaultmidpunct}
{\mcitedefaultendpunct}{\mcitedefaultseppunct}\relax
\EndOfBibitem
\bibitem[Wu \emph{et~al.}(2018)Wu, Pascal, Baskin, Wang, Fang, Liu, Lu, Guo, Prendergast, and Salmeron]{Wu2018}
C.~H. Wu, T.~A. Pascal, A.~Baskin, H.~Wang, H.~T. Fang, Y.~S. Liu, Y.~H. Lu, J.~Guo, D.~Prendergast and M.~B. Salmeron, \emph{Journal of the American Chemical Society}, 2018, \textbf{140}, 16237--16244\relax
\mciteBstWouldAddEndPuncttrue
\mciteSetBstMidEndSepPunct{\mcitedefaultmidpunct}
{\mcitedefaultendpunct}{\mcitedefaultseppunct}\relax
\EndOfBibitem
\bibitem[Sanz-Matìas \emph{et~al.}(2023)Sanz-Matìas, Roncoroni, Sundararaman, and Prendergast]{Sanz-Matias2023}
A.~Sanz-Matìas, F.~Roncoroni, S.~Sundararaman and D.~Prendergast, \emph{Ca-dimers and solvent layering determine electrochemically active species in \ce{Ca(BH4)2} in THF}, 2023, arXiv:2303.08261\relax
\mciteBstWouldAddEndPuncttrue
\mciteSetBstMidEndSepPunct{\mcitedefaultmidpunct}
{\mcitedefaultendpunct}{\mcitedefaultseppunct}\relax
\EndOfBibitem
\end{mcitethebibliography}
